\documentclass[referee]{raa}            

\usepackage{amssymb}
\usepackage{lipsum}

\usepackage{gensymb}
\usepackage{amsmath}
\usepackage{hyperref} 
\usepackage{epstopdf}
\usepackage{xcolor}
\usepackage{booktabs}
\usepackage{multirow}
\usepackage{graphicx, times}             
\usepackage{natbib}
\bibpunct{(}{)}{;}{a}{}{,}

\begin{document}

\title{Exploring the Structure and Evolution of Four Young Open Clusters Near the Galactic Mid-plane via Gaia DR3}

   \volnopage{Vol.0 (20xx) No.0, 000--000}      
   \setcounter{page}{1}          
\author{W. H. Elsanhoury
      \inst{1,*}\footnotetext{$*$Corresponding author, \it elsanhoury@nbu.edu.sa}
   \and S. Taşdemir
      \inst{2}
   \and D. C. Çınar
      \inst{2}
   \and R. Canbay
      \inst{3}
      \and A. A. Haroon
      \inst{4,5}
      \and A. Ahmed
      \inst{6}
}

\institute{Physics Department, College of Science, Northern Border University, Arar, Saudi Arabia\\
        \and
          Istanbul University, Institute of Graduate Studies in Science, Programme of Astronomy and Space Sciences, Beyaz{\i}t, Istanbul 34116, Turkey\\
        \and
          Istanbul University, Faculty of Science, Department of Astronomy and Space Sciences, Beyaz{\i}t, Istanbul 34116, Turkey\\
        \and
         Astronomy and Space Science Department, Faculty of Science, King Abdulaziz University, Jeddah, Saudi Arabia\\
        \and
          Astronomy Department, National Research Institute of Astronomy and Geophysics (NRIAG), Helwan 11421, Cairo, Egypt\\
        \and
           Astronomy, Space Science and Meteorology Department, Faculty of Science, Cairo University, Giza 12613, Cairo, Egypt\\
}
\vs\no

\abstract{We present a comprehensive analysis of four young open clusters, NGC 663, NGC 2301, NGC 2384, and NGC 7510, utilizing high-precision astrometric and photometric data from \textit{Gaia} DR3. Cluster membership was determined using the \texttt{UPMASK} algorithm, resulting in probable member counts ranging from 337 to 1498 across the clusters. Bayesian MCMC isochrone fitting yielded cluster ages in the range $\log t \sim$7.0–8.15, with uncertainties of $\sim$0.11–0.18. Reddening values ranged from $E(B-V) = 0.093$ mag in NGC 2301 to 1.24 mag in NGC 7510, consistent with their positions near the Galactic plane. The stellar mass function slopes ($\alpha \approx 2.00$--2.26) closely match the Salpeter IMF, with total stellar masses spanning nearly an order of magnitude, from $\sim$486$\,M_\odot$ in NGC 2301 to $\sim$3584$\,M_\odot$ in NGC 663. Dynamical relaxation times indicate that only NGC 2301 ($\tau \approx 10.00$) and NGC 2384b ($\tau \approx 2.03$) are dynamically relaxed; the others remain dynamically evolving. Orbital integrations in the MWPotential2014 model reveal nearly circular Galactic orbits with very low eccentricities ($e \approx 0.003$--0.014) and small vertical distances ($Z_{\rm max} < 0.142$ kpc), confirming their confinement to the thin disk. SED and kinematic analysis show that NGC 2384 a \& b are separated by $\sim$0.55 kpc, indicating an optical pairs.}


\authorrunning{Elsanhoury et al.}            
\titlerunning{Exploring the Structure and Evolution of Four Young OCs}  %
\maketitle

\section{Introduction}\label{Introduction}

Stars from the same chemical cloud that are gravitationally bound and formed together are known as open clusters (OCs). These clusters, which are mostly found in the Galactic disk, particularly close to the Galactic plane, where interstellar gas and dust densities are maximum, are important indicators of recent star formation.  Since there are many young stellar populations due to the concentration of star-forming material in this area, OCs are essential resources for understanding the composition and development of the Milky Way (MW).

The mid-plane usually lies a few hundred parsec (pc) away from OCs close to the Galactic plane \citep{2016A&A...593A.116J}. Compared to globular clusters, they are younger and less securely linked, and dynamical processes like differential Galactic rotation and tidal interactions with molecular clouds limit their lifespans. This causes a large number of OCs to vanish within a few hundred million years into the Galactic field population \citep{2018A&A...614A..22P}. Despite their significance, OCs close to the Galactic plane are difficult to observe. Interstellar dust causes substantial extinction in the region, and field star contamination complicates the determination of cluster membership. The identification, characterization, and dynamical study of OCs, even in very crowded and obscured fields, have been transformed by recent developments in space-based astrometry, particularly data from the \textit{Gaia} mission \citep{Cantat_2020, CastroGinard2022}.

\subsection{Target Clusters and Literature Review}

\subsubsection{NGC 663}
NGC 663 is a moderately rich OC located in the Perseus arm of MW. It has been extensively studied because of its young age and relatively high stellar density. Several photometric analyses have estimated that its age is in the range of 20--25 Myr (log~$t \approx 7.3$--$7.4$), indicating that the cluster is still in the early stages of stellar evolution \citep{pigulski2001,Fabregat2005}. Its heliocentric distance has been reported as approximately 2.1 kpc, corresponding to a distance modulus of around 11.6 mag. The reddening across the cluster is spatially variable, with typical values of colour excess reaching up to $E(R-I) \approx 0.54$ mag. Despite the availability of detailed photometric and astrometric data, metallicity estimates for this cluster remain sparse in the literature.

\subsubsection{NGC 2301}

NGC 2301 is an intermediate-age open cluster located in the Galactic anti-centre direction. It has been a subject of interest due to its low reddening and well-defined main sequence (MS). Its age is estimated to be between 165 and 200 Myr, making it one of the older clusters considered in this study \citep{Sukhbold2009}. The cluster is relatively nearby, with a heliocentric distance of around 870 pc. The reddening is minimal, with $E(B-V) \approx 0.04$ mag. Photometric and spectroscopic studies generally report a metallicity close to the solar value ([Fe/H]~$\approx$~0.0 dex). The cluster’s favourable observational conditions make it an excellent candidate for comparative isochrone and SED-based parameter estimation.

\subsubsection{NGC 2384}

NGC 2384 is a young open cluster situated in the Monoceros constellation. Its age has been estimated to be within the range of 12 to 20 Myr, placing it among the youngest objects in our sample \citep{hasan2008,Vazquez2010}. The cluster lies at a distance of approximately 3.0--3.2 kpc from the Sun. The interstellar extinction toward the cluster is moderate, with a reported colour excess of $E(B-V) \approx 0.31$ mag. Although it has been included in several photometric surveys, detailed metallicity measurements are lacking, and the cluster is generally assumed to possess a near-solar composition. Due to its youth and location near the Galactic plane, NGC 2384 provides valuable insights into early-stage cluster dynamics and Galactic structure.

\subsubsection{NGC 7510}
Located in the direction of the Cepheus constellation, NGC 7510 is a compact, young open cluster characterized by high extinction and differential reddening. Previous investigations have determined that its age lies between 10 and 18 Myr, consistent with its classification as a young stellar population \citep{sagar1991,1996A&AS..115..325B}. The cluster is situated at a heliocentric distance of about 2.8 to 3.2 kpc, based on photometric distance modulus estimations. Its average colour excess is notably high, ranging from $E(B-V) \approx 1.0$ to $1.3$ mag. The cluster is generally assumed to have near-solar metallicity, although high-resolution spectroscopic measurements are limited. More recently, studies based on \textit{Gaia} have refined its structural and 
kinematic parameters.
\begin{figure}[ht]
\centering
\includegraphics[width=0.8\linewidth]{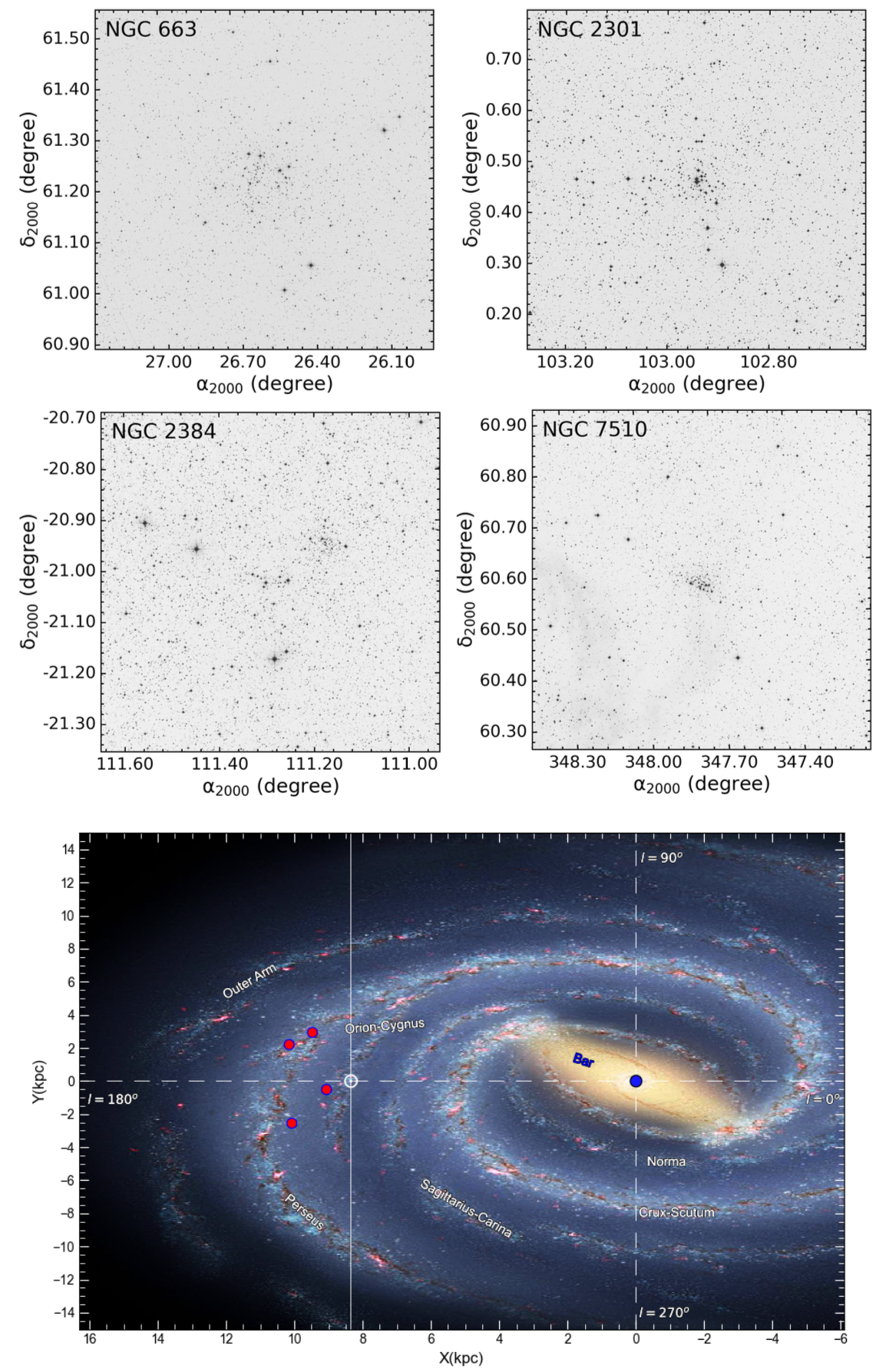}
\caption{Digitized Sky Survey (DSS) images displaying the identification of the open clusters NGC 663, NGC 2301, NGC 2384, and NGC 7510 with the position of each cluster on the Milkyway.}
\label{fig: DSS}
\end{figure}

The current work presents a detailed photometric, spectroscopic, and dynamical study of four open star clusters, NGC 663, NGC 2301, NGC 2384, and NGC 7510, using the most recent data of astrometric and photometric data from the $Gaia$ DR3 database \citep{GaiaDR3}. Figure \ref{fig: DSS} shows their Digitized Sky Survey (DSS) images. For illustrative purposes, the surrounding stellar fields of these clusters were extracted from the DSS archive by employing the \texttt{POSS2/UKSTU Red} photographic plates\footnote{\url{https://archive.stsci.edu/cgi-bin/dss_form}}, providing a direct view of their projected positions against the Galactic background.

This paper is organized as follows: The details of the data used in this study are presented in Section \ref{sec:Data}. Structural parameters belong to the each clusters are given in section \ref{sec:Structural}. We discuss the results of the $Gaia$ DR3 data analysis of the selected OCs, including the estimation of their astrometric (\ref{sec:astrometric}), photometric ( \ref{sec:astrophysics}),  spectral energy (\ref{sec:SED}) distribution (SED), mass and luminosity functions (\ref{sec:LF and MF}), kinematic parameters 
(\ref{sec:kinematics}), dynamical (\ref{sec:dynamic and kinematic}), and orbital parameters in section  \ref{sec:orbit}. Finally, in Section \ref{Summary and conclusion} we present our conclusion about the main result.

\section{Data}
\label{sec:Data}

The European Space Agency’s \textit{Gaia} mission has revolutionized stellar astrophysics by providing high-precision astrometric, photometric, and spectroscopic measurements for more than 1.8 billion sources. The third data release (\textit{Gaia} DR3; \citealt{GaiaDR3}) provides five-parameter astrometry ($\alpha$, $\delta$, $\varpi$, $\mu_{\alpha}\cos\delta$, $\mu_{\delta}$), broadband photometry ($G$, $G_{\rm BP}$, $G_{\rm RP}$), and radial velocities ($V_{\rm r}$) for a significant fraction of stars, offering unprecedented opportunities for detailed studies of Galactic open clusters (OCs). 

For the present work, we compiled catalogues for four open clusters: NGC 663, NGC 2301, NGC 2384, and NGC 7510. The central equatorial coordinates of each cluster were adopted from \citet{Cantat_2020}. A 40 arcmin region around the centers of NGC 663, NGC 2384, and NGC 7510, and a 60 arcmin region for NGC 2301 were extracted to encompass both the cluster cores and their extended coronae. These regions include 100,361 stars in NGC 663, 179,625 stars in NGC 2301, 130,137 stars in NGC 2384, and 88,485 stars in NGC 7510. The initial catalogues span the magnitude range $7 < G \leq 22$ mag.

The bottom panel of Figure~\ref{fig: DSS} also displays the locations of the selected open clusters on the Galactic plane, emphasizing their distribution along the spiral arms and near the midplane. The extracted 40–60 arcmin regions around the cluster centres encompass dense stellar fields. Studying clusters in these crowded midplane regions is crucial, as they trace the young stellar population of the Galactic disk and provide key insights into its structure, kinematics, and chemical evolution. Their spatial and dynamical properties offer valuable constraints on the formation and evolution of the Galactic disk.

\subsection{Photometric Completeness Assessment}

\begin{figure}[h]
\centering
\includegraphics[width=0.95\linewidth]{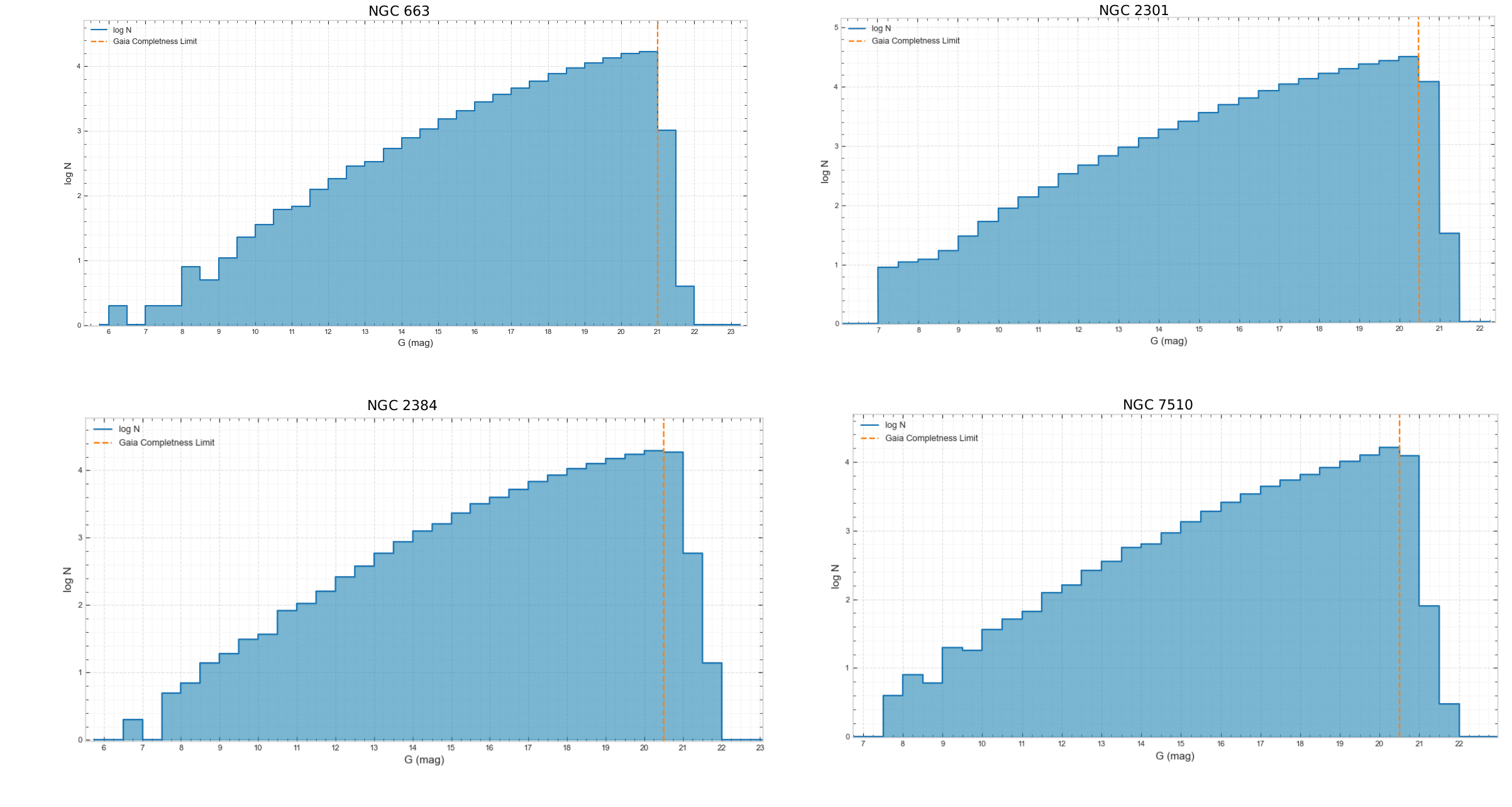}
\caption{Histograms of \textit{Gaia} photometric data for the OCs NGC 663, NGC 2301, NGC 2384, and NGC 7510. The orange dashed line indicates the \textit{Gaia} completeness limit in each OC.}
\label{fig:completness}
\end{figure}

Determining the photometric completeness limit is a prerequisite for robust parameter estimation of OCs. To evaluate the limits, we constructed $G$-band magnitude histograms for each field. As expected, the number counts rise steadily toward fainter magnitudes before turning over due to incompleteness.
The turnover points were adopted as the limiting magnitudes: $G=21.0$ mag for NGC 663 and $G=20.50$ mag for NGC 2301, NGC 2384, and NGC 7510 (see Figure~\ref{fig:completness}). Stars fainter than these limits were statistically excluded from the analysis to minimize biases in the derived astrophysical. After applying the photometric completeness limit for each cluster, the number of stars decreased to 91378, 152899, 100639 and 67515 for NGC 663, NGC 2301, NGC 2384 and NGC 7510, respectively.

\subsection{Uncertainties in Gaia DR3 Measurements}

The uncertainties in the \textit{Gaia} DR3 photometry and astrometry were taken as functions of $G$ magnitude. For bright stars ($G<15$ mag), typical uncertainties are $\sim0.002$–0.003 mag in $G$, $\sim0.02$–0.03 mas in trigonometric parallax, and $\sim0.02$–0.03 mas~yr$^{-1}$ in proper motions. At $G\simeq20$ mag, these errors increase to $\sim0.01$ mag in $G$, $\sim0.5$ mas in trigonometric parallax, and $\sim0.5$ mas~yr$^{-1}$ in proper motions, while at the faint end ($G\approx21$ mag) the uncertainties can reach $\sim0.02$ mag, $\sim1.3$ mas, and $\sim1.4$ mas~yr$^{-1}$, respectively \citep{GaiaDR3}. These error characteristics were explicitly considered in subsequent membership determination and isochrone fitting procedures.
The final cleaned catalogues thus consist of high-quality astrometric and photometric measurements for all stars brighter than the adopted completeness limits in each cluster region. The corresponding photometric error distributions per magnitude bin are presented in Table~\ref{tab:photometric_errors}.

\begin{table*}[h]
\centering
\caption{Errors in apparent magnitudes ($G$) and colours ($\sigma_{G_{\rm BP}-G_{\rm RP}}$) of stars located along the direction of the selected open clusters.}
\label{tab:photometric_errors}
\resizebox{\textwidth}{!}{%
\begin{tabular}{c|ccc|ccc|ccc|ccc}
\toprule
 & \multicolumn{3}{c|}{\textbf{NGC 663}} & \multicolumn{3}{c|}{\textbf{NGC 2301}} & \multicolumn{3}{c|}{\textbf{NGC 2384}} & \multicolumn{3}{c}{\textbf{NGC 7510}} \\
\cmidrule(lr){2-4} \cmidrule(lr){5-7} \cmidrule(lr){8-10} \cmidrule(lr){11-13} $G$ (mag) & $N$ & $\sigma_{\rm G}$ & $\sigma_{G_{\rm BP}-G_{\rm RP}}$ & $N$ & $\sigma_{\rm G}$ & $\sigma_{G_{\rm BP}-G_{\rm RP}}$ & $N$ & $\sigma_{\rm G}$ & $\sigma_{G_{\rm BP}-G_{\rm RP}}$ & $N$ & $\sigma_{\rm G}$ & $\sigma_{G_{\rm BP}-G_{\rm RP}}$ \\
\hline \hline
6--14   & 1383  & 0.0028 & 0.0055 & 3550  & 0.0029 & 0.0059 & 2097  & 0.0029 & 0.0057 & 1380  & 0.0029 & 0.0059 \\
14--15  & 1547  & 0.0028 & 0.0061 & 3663  & 0.0029 & 0.0059 & 2483  & 0.0029 & 0.0054 & 1370  & 0.0029 & 0.0057 \\
15--16  & 3068  & 0.0028 & 0.0060 & 7091  & 0.0030 & 0.0073 & 4673  & 0.0029 & 0.0066 & 2717  & 0.0029 & 0.0066 \\
16--17  & 5692  & 0.0029 & 0.0083 & 12571 & 0.0031 & 0.0117 & 8103  & 0.0029 & 0.0089 & 5243  & 0.0029 & 0.0099 \\
17--18  & 9399  & 0.0029 & 0.0149 & 20974 & 0.0033 & 0.0229 & 13614 & 0.0030 & 0.0152 & 8901  & 0.0031 & 0.0207 \\
18--19  & 15301 & 0.0032 & 0.0313 & 31769 & 0.0039 & 0.0527 & 21152 & 0.0033 & 0.0308 & 13186 & 0.0035 & 0.0472 \\
19--20  & 22773 & 0.0038 & 0.0655 & 44864 & 0.0058 & 0.1178 & 29988 & 0.0042 & 0.0669 & 20399 & 0.0046 & 0.0980 \\
20--21  & 32214 & 0.0075 & 0.1523 & 54089 & 0.0130 & 0.2672 & 39617 & 0.0082 & 0.1655 & 32145 & 0.0105 & 0.2243 \\
21--23  & 8983  & 0.0242 & 0.3938 & 1053  & 0.0296 & 0.4507 & 8409  & 0.0230 & 0.4008 & 3143  & 0.0255 & 0.4269 \\
\bottomrule
\end{tabular}}
\end{table*}

\section{Structural Analysis}
\label{sec:Structural}
\subsection{Redetermination of the OCs Centres}

A straightforward yet effective technique that works especially well with big datasets like \textit{Gaia} DR3 is the re-determination of OCs centres using histograms. Histograms give an approachable and straightforward method that is particularly helpful in preliminary analyses and instructional contexts, even while more complex methods such as Kernel Density Estimation (KDE) or machine learning models offer more precision. The histogram approach is crucial for enhancing our comprehension of OCs formation and dynamics.

\begin{figure}[h]
\centering
\includegraphics[width=0.75\linewidth]{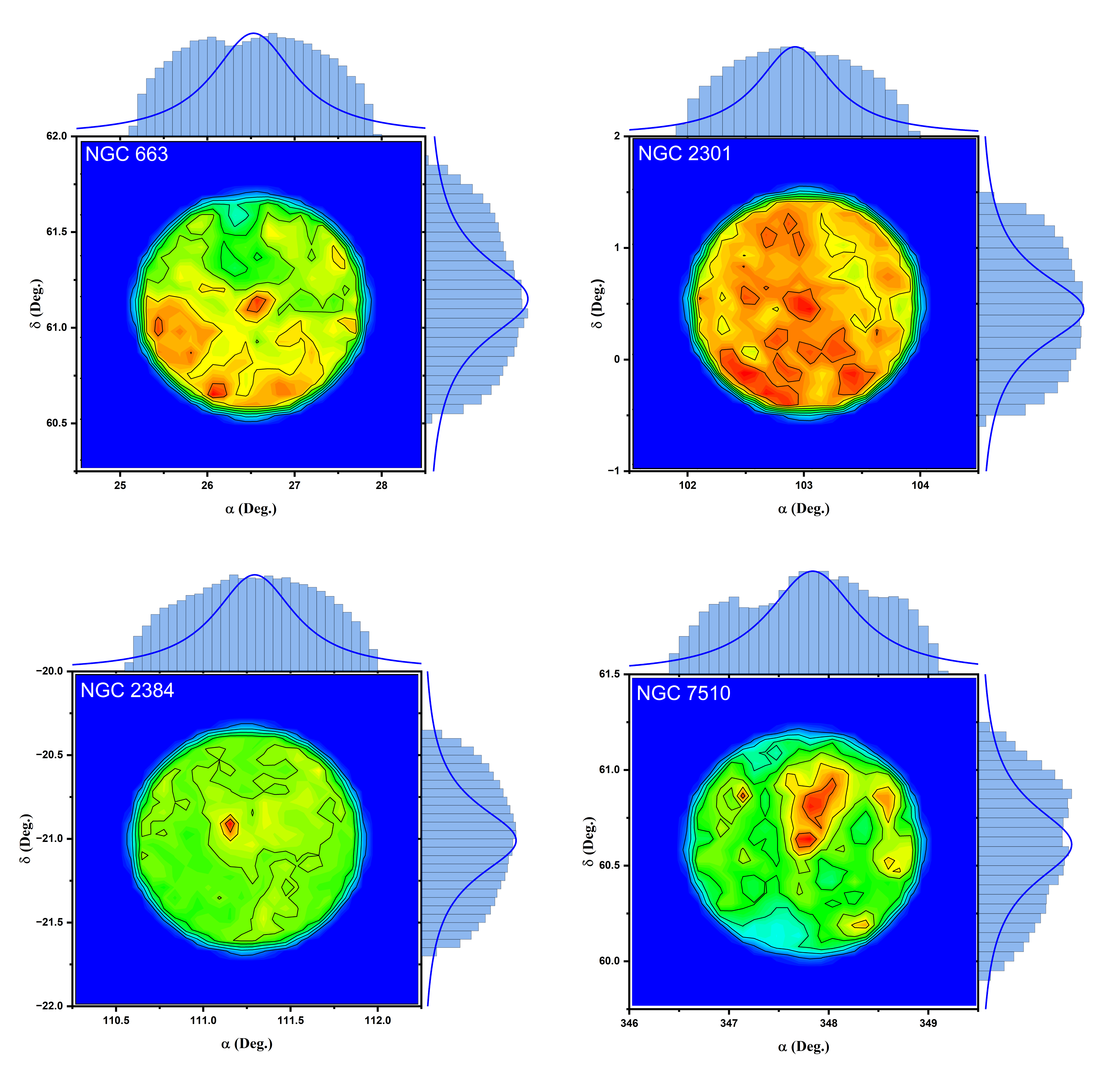}
\caption{The Gaussian fit provides the coordinates of the highest density areas density areas in $\alpha$ and $\delta$, as well as the density contour diagrams illustrating the structural properties of of NGC 663, NGC 2301, NGC 2384, and NGC 7510.}
\label{fig: centers}
\end{figure}

By identifying the regions with the highest star density, the cluster cores can be identified. To achieve this, we created $\alpha$ and $\delta$ histograms, separated the extracted region into identically sized bins, and used Gaussian fitting. Therefore, the clusters centres exit at $(\alpha,~\delta~\&~l,~b)_{NGC~663}$ = (01$^h$ 46$^m$ 06$^s$.96, 61$^{\circ}$ 09$'$ 03$''$.60 $\&$ 129$^{\circ}$.835, -0$^{\circ}$.691), $(\alpha,~\delta~\&~l,~b)_{NGC~2301}$ = (06$^h$ 51$^m$ 41$^s$.76, 00$^{\circ}$ 26$'$ 49$''$.20 $\&$ 212$^{\circ}$.913, 0$^{\circ}$.804), $(\alpha,~\delta~\&~l,~b)_{NGC~2384}$ = (07$^h$ 25$^m$ 10$^s$.80, -21$^{\circ}$ 00$'$ 32$''$.40 $\&$ 235$^{\circ}$.710, -1$^{\circ}$.988), and $(\alpha,~\delta~\&~l,~b)_{NGC~7510}$ = (23$^h$ 11$^m$ 21$^s$.60, 60$^{\circ}$ 36$'$ 39$''$.60 $\&$ 111$^{\circ}$.299, 0$^{\circ}$.243) as seen in Figure \ref{fig: centers}.

\subsection{Radial Density Profile}
\begin{figure}[ht]
\centering
\includegraphics[width=0.75\linewidth]{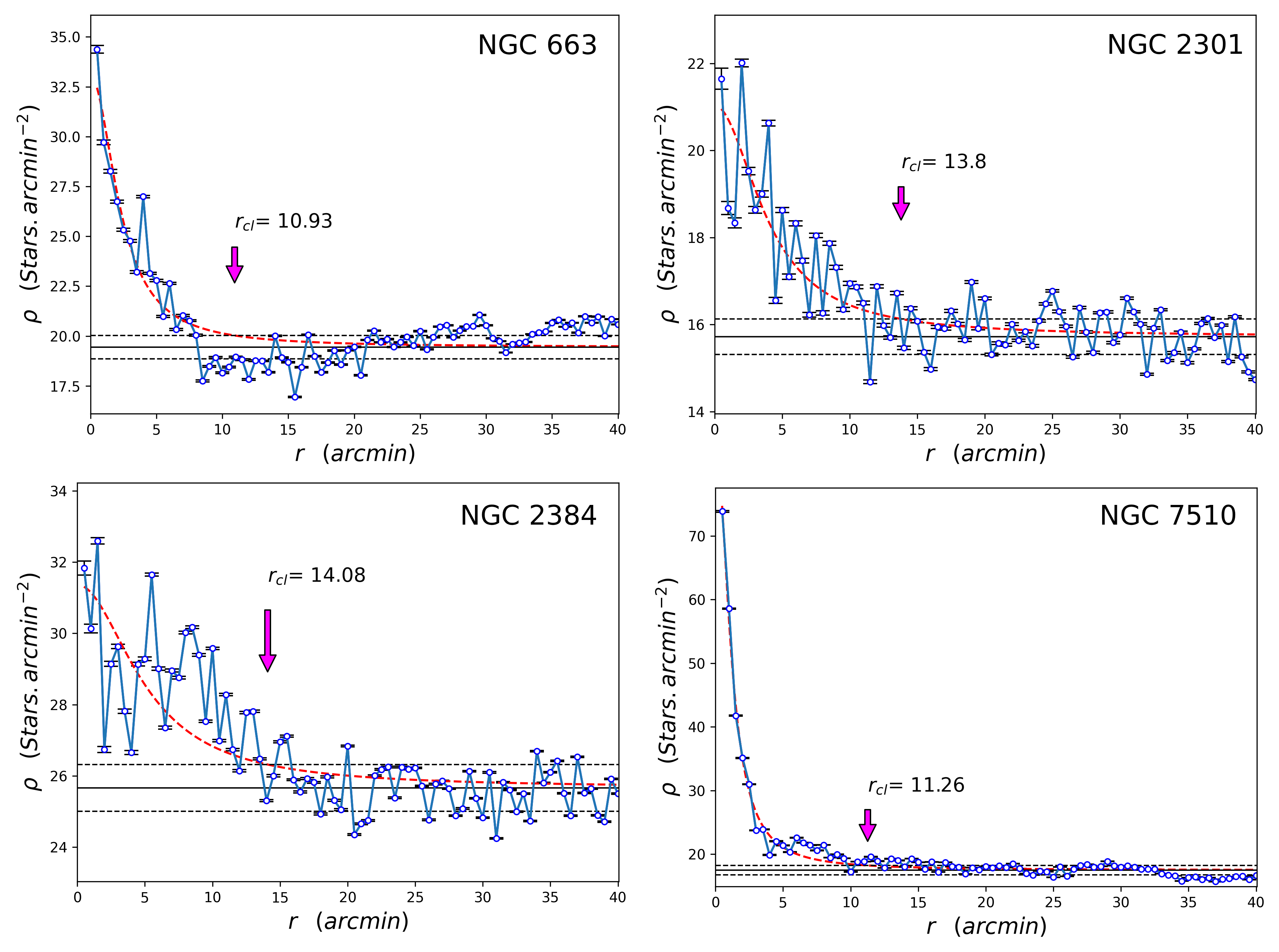}
\caption{RDPs of NGC 663, NGC 2301, NGC 2384, and NGC 7510 OCs. The curved solid lines represent the fitting of King's model \citep{king1966} and the dashed lines represent the background field density $f_{bg}$.}
\label{fig: RDP}
\end{figure}

To study the structure of the OCs, we drew the radial density profile (RDP), as shown in Figure \ref{fig: RDP}, by dividing the observed area into concentric shells from the cluster centre with equal steps of 0.50 arcmin. Each zone's density ($\rho_i$; stars arcmin\textsuperscript{-1})  was computed by dividing the number of stars within it by its area, i.e. $\rho_i=\frac{N_i}{A_i}$.
The uncertainties in the density calculations were determined from sampling statistics, in accordance with Poisson distribution ($= 1/\sqrt{N_i}$).

\begin{table*}
\centering
\caption{The structural parameters of the clusters derived from King profile fitting, including the core radius ($ r_c$), core limiting radius ($r_{cl}$), central star ($\rho_0$) and background stellar densities ($\rho_{bg}$),  concentration parameters ($C$), and density contrast parameter ($\delta_c$).
\label{table-king}}
\begin{tabular}{lcccc}
\hline
\textbf{Parameters} & \textbf{NGC 663}  & \textbf{NGC 2301} & \textbf{NGC  2384}  & \textbf{NGC 7510} \\
\hline
\hline
$ r_c$ (arcmin)  &  2.32 $\pm$ 0.06 & 3.98 $\pm$ 0.05 & 5.08 $\pm$ 0.04 & 1.19 $\pm$ 0.09\\
$ r_c$ (pc)  & 1.95 $\pm$ 0.07 & 1.02 $\pm$ 0.01 & 4.65 $\pm$ 0.05 & 1.10 $\pm$ 0.01\\
$r_{cl}$  (arcmin) & 10.93 $\pm$ 3.31 & 13.80 $\pm$ 3.71 & 14.08 $\pm$ 3.75 & 11.26 $\pm$ 3.56\\
$r_{cl}$  (pc) & 9.19 $\pm$ 3.03 & 3.55 $\pm$ 1.88 & 12.89 $\pm$ 3.60 & 10.33 $\pm$ 3.21\\
$\rho_0$ ($\rm stars.arcmin^{-2}$)  &13.59 $\pm$ 3.69 & 5.31 $\pm$ 2.30& 5.70 $\pm$ 2.39 & 66.94 $\pm$ 8.18\\
$\rho_{bg}$ $\rm (stars.arcmin^{-2})$ & 19.44 $\pm$ 4.41 & 15.72 $\pm$ 3.97 & 25.66 $\pm$ 5.07 & 17.48 $\pm$ 4.18 \\
$C$& 4.71 $\pm$ 0.46 & 3.47 $\pm$ 0.54 & 2.77 $\pm$ 0.60 & 9.46 $\pm$ 0.33 \\
$\delta_c$& 1.70 & 1.34 & 1.22 & 4.83 \\
\hline
\end{tabular}
\end{table*} 
The radius of the cluster is estimated as the distance at which the star density equals that of the field star density \citep{Tadross2005}. Applying the empirical King’s Model \citep{King1962}, the density function $\rho_r$ can be represented as:
\begin{equation}
\rho (r)= \rho_{bg} + \frac{\rho_o}{1+ (r/r_c)^2}\,, 
\end{equation}
where $\rho_{bg}$, $\rho_o$ and $r_c$ are background density, central star density, and the core radius (which is the distance where the stellar density drops to half of the central density) of the cluster, respectively. The best-fit model is the model with the highest coefficient of determination:
\begin{equation}
R^2=1-\frac{SS_{res}}{SS_{tot}}\,,
\end{equation}
$SS_{res}$ is the residual sum of squares computed by $SS_{res}=\sum_i(y_i-f_i)^2$ where $y_i$ represents the observed RDP, $f_i$ represents the King's profile, and $SS_{tot}$ is the total sum of squares computed by $SS_{tot}=\sum_i(y_i-\overline{y})^2$, where $\overline{y}$ is the mean of the observed data, $\overline{y}=\frac{1}{n}\times\sum_{i=1}^n y_i$.

The density contrast parameter ($\delta_{\rm c}$) and the concentration parameter ($C$) were calculated in order to further characterize the structural features of these clusters.  The density contrast measures the cluster's relative density to the surrounding stellar field and is expressed as $\delta_{\rm c} = 1 + \rho_o / \rho_{\rm bg}$.  Densely stored clusters tend to have higher degrees of central concentration, as indicated by higher values of $\delta_{\rm c}$, within the range $7 \lesssim \delta_{\rm c} \lesssim 23$. The core region's compactness is measured by the concentration parameter, which is defined as $C = (r_{\rm cl}/r_{\rm c})$ \citep{king1966}. $C = 4.71\pm0.46,~3.47\pm0.54,~2.77\pm0.60, 9.46\pm0.33$ for NGC 663, NGC 2301, NGC 2384, and NGC 7510, respectively, were the results of our analysis are shown in Table \ref{table-king}.
 
\section{The Photometric Analysis}

\subsection{Membership Determination}
\label{section:membership}

Cluster membership was established using the unsupervised photometric membership assignment algorithm \texttt{UPMASK} \citep{Krone-Martins_2014}, a robust, model-independent technique that statistically separates genuine cluster members from the surrounding field-star population. This algorithm combines $k$-means clustering with significance testing in a multi-dimensional parameter space, iteratively refining membership probabilities without imposing any prior assumptions about the cluster’s spatial or kinematic profile \citep{Yontan2022, TasdemirCinar_2025}.

For each cluster, we constructed a five-dimensional parameter space comprising equatorial coordinates ($\alpha, \delta$), trigonometric parallax ($\varpi$), and proper-motion components ($\mu_{\alpha}\cos\delta, \mu_\delta$), including their associated uncertainties. The membership analysis was executed over 25 outer-loop iterations per cluster to ensure convergence and robustness, and the optimal $k$-means clustering parameters were determined as 24, 20, 28, and 7 for the respective clusters.

\begin{figure}[h]
\centering
\includegraphics[width=1\linewidth]{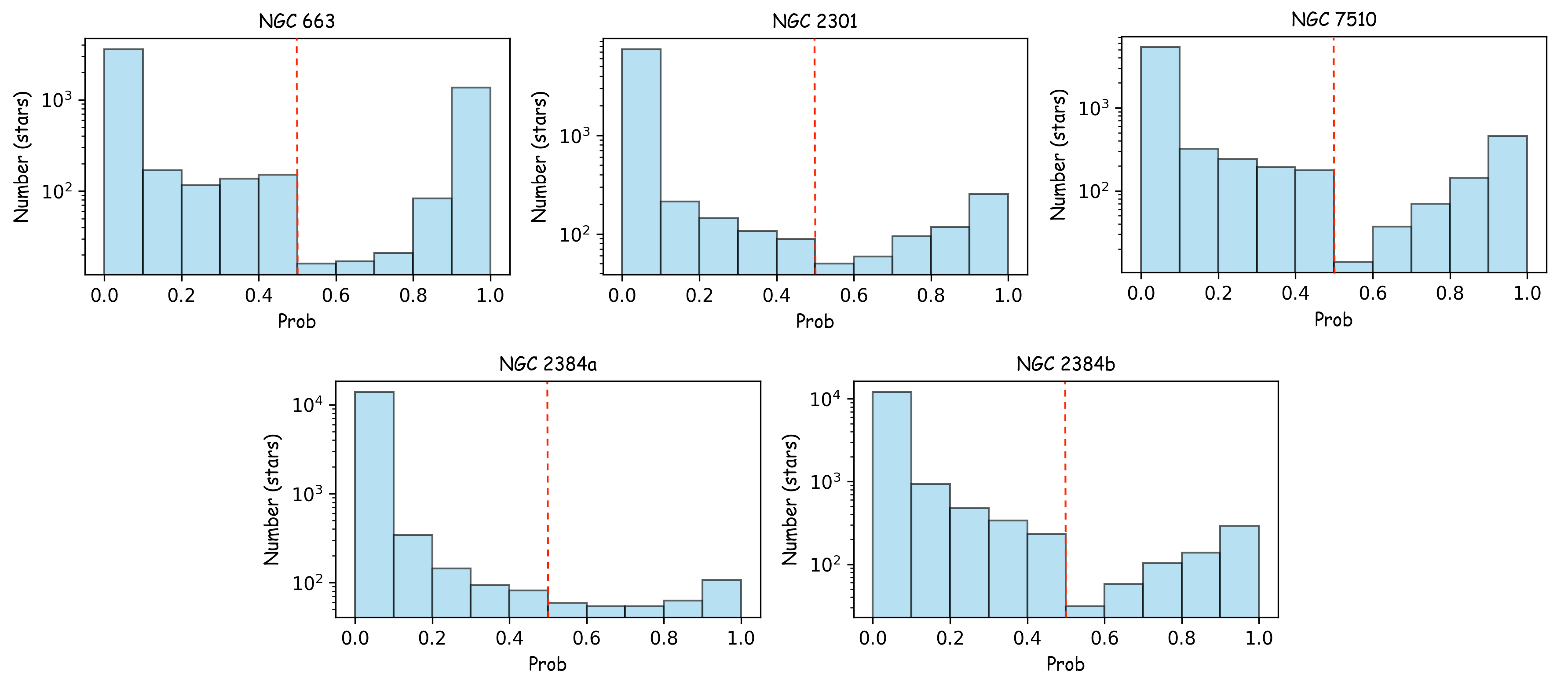}
\caption{Histogram of membership probabilities ($P$) from \texttt{UPMASK} for all stars in the studied clusters. The vertical red dashed line indicates the adopted membership probability threshold of $P \geq 50\%$, which separates likely cluster members from field stars.}
\label{fig:prob}
\end{figure}

Stars with membership probabilities $P \geq 50\%$ were selected as cluster members. The computed membership probabilities for all stars are presented in Figure~\ref{fig:prob}. Inspection of these distributions reveals a clear separation between high-probability stars, which are likely genuine cluster members, and low-probability stars, likely field contaminants. The adopted 0.5 threshold corresponds to the midpoint of this distribution, providing a natural and statistically motivated division that maximizes the inclusion of true members while minimizing contamination. This choice aligns with standard practices in the literature and ensures an optimal balance between completeness and purity. The reliability and robustness of the membership selection were further quantified following the methodology of \cite{1996AcASn..37..377S}, confirming the effectiveness of our procedure in distinguishing cluster members from field stars.

\begin{table*}[ht]
\centering
\renewcommand{\arraystretch}{0.75}
\footnotesize
\caption{Comparison of the estimated parameters in the literature for OCs:($\alpha$, $ \delta$), radius ($r$), age ($t$), distance ($d$), colour excess ($E(B - V)$), trigonometric parallax ($\varpi$), proper motion components ($\mu_\alpha \cos \delta, \mu_\delta$), the number of member stars (N) and the references.
\label{tab:literature}}    
\setlength\tabcolsep{0.8pt}
\begin{tabular}{cccccccccccc}
\toprule
$\alpha $ &  $\delta$  &  $r$ & $t$  & $d$ & \it E(B-V) & $ \varpi$ &  $ \mu _\alpha \cos \delta$ & $ \mu _\delta$  & N & Ref.  \\
\ (deg) & (deg) &  (arcmin) & (Myr) & (kpc) & (mag) & (mas) &  (mas $\rm yr^{-1}$) & (mas $\rm yr^{-1}$)& stars & \\
\hline\hline
\multicolumn{11}{c}{\textbf{NGC 663}}\\
\midrule
26.586  & 61.212  & 9.0$^*$   & 29.5 & 2.950 & --& 0.343$\pm$0.025 & $-$1.141$\pm$0.064 &  $-$0.319$\pm$0.069& 455&1 \\
26.586  & 61.212  & 9.0$^*$ & -- & 2.865 &--& 0.320$\pm$0.002&$-$1.11$\pm$0.003& $-$0.235$\pm$0.004 & 623 & 2 \\
26.538 &	61.235 & 8.67 &  25.1 & 2.420 &0.8&--& $-$2.56	$\pm$4.05  &$-$1.28$\pm$3.27& 444 &3 \\
26.538 & 61.235 & 8.0  & -- & -- & -- &--& $-$2.52$\pm$1.27 & $-$0.96 $\pm$0.67
 & 518 &4\\
26.550 & 61.225 & 14.4  & 31.6 & 2.100 & 0.7  &--& $-$0.98 & $-$1.94  & -- &5 \\
\hline\hline
\multicolumn{11}{c}{\textbf{NGC 2301}}\\
\midrule
102.943  & 0.465  & 9.72$^*$   & 213.8 & 0.8570 & --& 1.150$\pm$0.035 & $-$1.358$\pm$0.158 &  $-$2.180$\pm$0.149& 399 &1 \\
102.943 & 0.465  & 9.72$^*$ & -- & 0.8595 &--& 1.135$\pm$0.004&$-$1.367$\pm$0.010 & $-$2.179$\pm$0.009 & 455 & 2 \\
102.938 & 	0.460 & 8.67 &  158.5 & 0.870 &0.03 &--& $-$0.23	$\pm$4.85  &$-$3.26$\pm$4.56& 111 &3 \\
102.938  & 0.460 & 8.0  & -- & -- & -- &--& $-$0.83$\pm$0.63 & $-$2.76$\pm$0.50
 & 355 &4 \\
102.945 & 0.465 & 15.6  & 223.9 & 0.805 & 0.062   &--& $-$0.71 & $-$4.07  & -- &5 \\

102.945 & 0.465 & -- & 223.9 & 0.805 & 0.062 &--& $-$0.83$\pm$0.63 & $-$2.76$\pm$0.50
 & 355 &6 \\
\hline\hline
\multicolumn{11}{c}{\textbf{NGC 2384}}\\
\midrule
111.292 & $-$21.022  & 3.0   & 14.1 & 3.070 &0.31& -- & $-$1.33$\pm$4.82 &  1.71$\pm$5.18& 55 &3 \\
111.292 & $-$21.022 & 3.5  & -- & -- & -- &--& $-$1.29$\pm$2.08 & 1.37 $\pm$4.65
 & 106 &4 \\
111.285 & $-$21.020 &9.9  & 13.5 & 2.074 & 0.396 &--& $-$6.44& 3.29  & -- &5 \\
\hline\hline
\multicolumn{11}{c}{\textbf{NGC 7510}}\\
\midrule
347.767 & 60.579  & 3.3$^*$   & 19.5 & 2.931 & --& 0.308$\pm$0.031 & $-$3.654$\pm$0.102 &  $-$2.235$\pm$0.085& 289 &1 \\
347.767 &  60.579  & 3.3$^*$ & -- & 3.178 &--& 0.286$\pm$0.003&$-$3.664$\pm$0.008& $-$2.193$\pm$0.008 & 324 & 2 \\
347.763 &	60.570 & 3.41 &  22.4 & 3.480 &0.9&--& $-$2.96$\pm$5.37  & 0.00$\pm$4.2& 100 &3 \\
347.762 & 60.570 & 4.0 &  -- & -- &--&--& $-$3.59$\pm$1.36  &$-$0.86$\pm$0.10& 136 &4\\
347.757 & 60.579  & 8.0  & 50.1 & 2.100 & 0.949  &--& $-$2.83 & $-$1.46  & -- &5 \\
\bottomrule
\end{tabular}
\newline
(1)~\cite{Poggio_2021}; (2)~\cite{Cantat_2020}; (3)~\cite{Sampedro2017}; (4)~\cite{Dias2014}; (5)~\cite{Kharchenko2016}; (6) ~\cite{2016A&A...593A.116J}. 
\\$^*$ These values are $r_{50}$ or the radii containing half the members.
\end{table*}

\subsection{Astrometric Properties}
\label{sec:astrometric}

\begin{figure}[ht]
\centering
\includegraphics[width=1\linewidth]{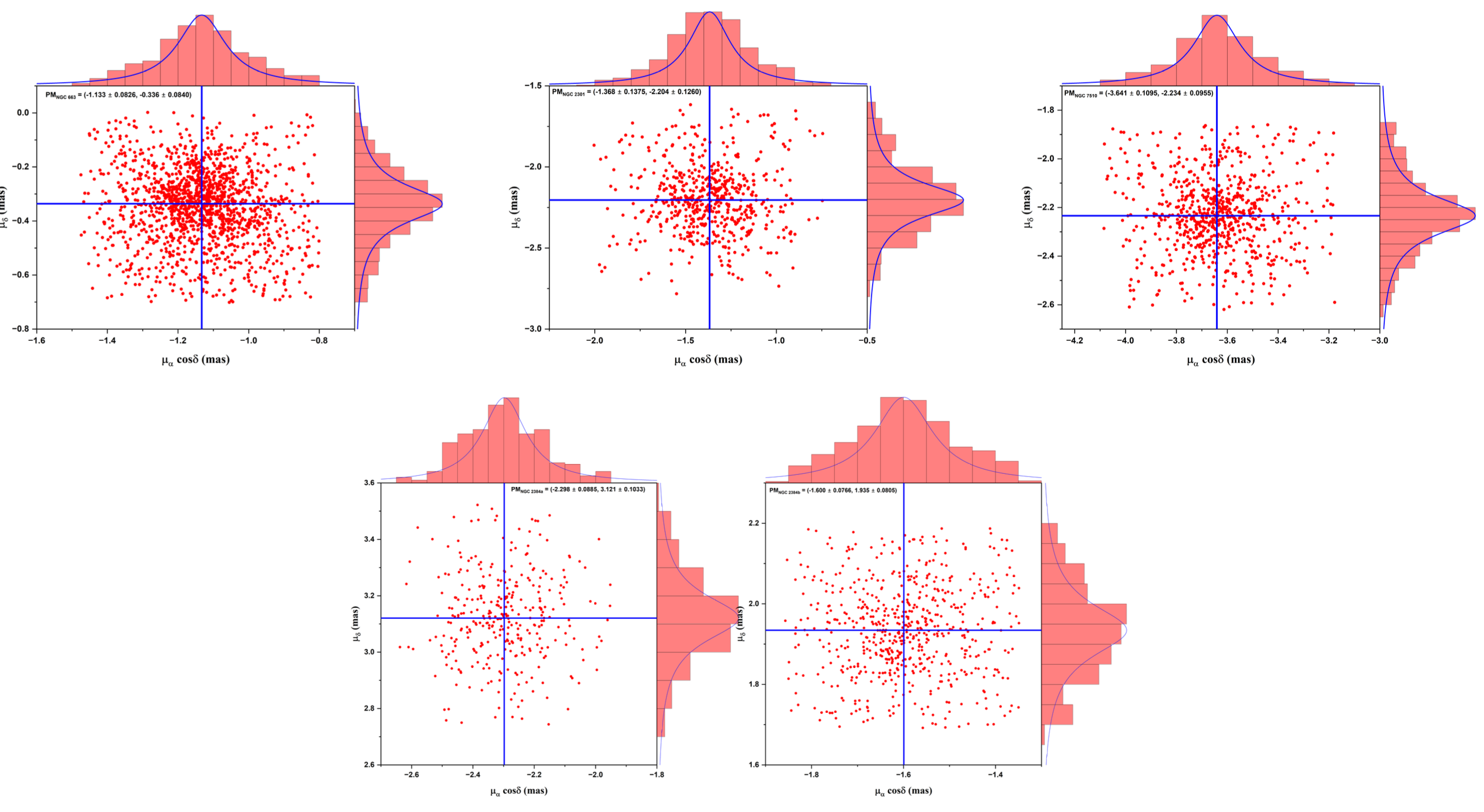}

\caption{The proper-motion components range in equatorial coordinates. The Gaussian fitting of the bins represented by blue lines in the panels. 
\label{fig: PM}}
\end{figure}

The mean proper-motion components in both directions of NGC 663 (-1.133 $\pm$ 0.0826, -0.336 $\pm$ 0.0840; mas yr$^{-1}$), NGC 2301 (-1.368 $\pm$ 0.1375, -2.204 $\pm$ 0.1260; mas yr$^{-1}$), NGC 2384a (-2.298 $\pm$ 0.0885, 3.121 $\pm$ 0.1033; mas yr$^{-1}$), NGC 2384b (-1.600 $\pm$ 0.0766, 1.935 $\pm$ 0.0805; mas yr$^{-1}$), and NGC 7510 (-3.641 $\pm$ 0.1095, -2.234 $\pm$ 0.0955; mas yr$^{-1}$), where the created and the constructed histograms are devoted with Figure \ref{fig: PM}.

\begin{figure}[ht]
\centering
\includegraphics[width=0.95\linewidth]{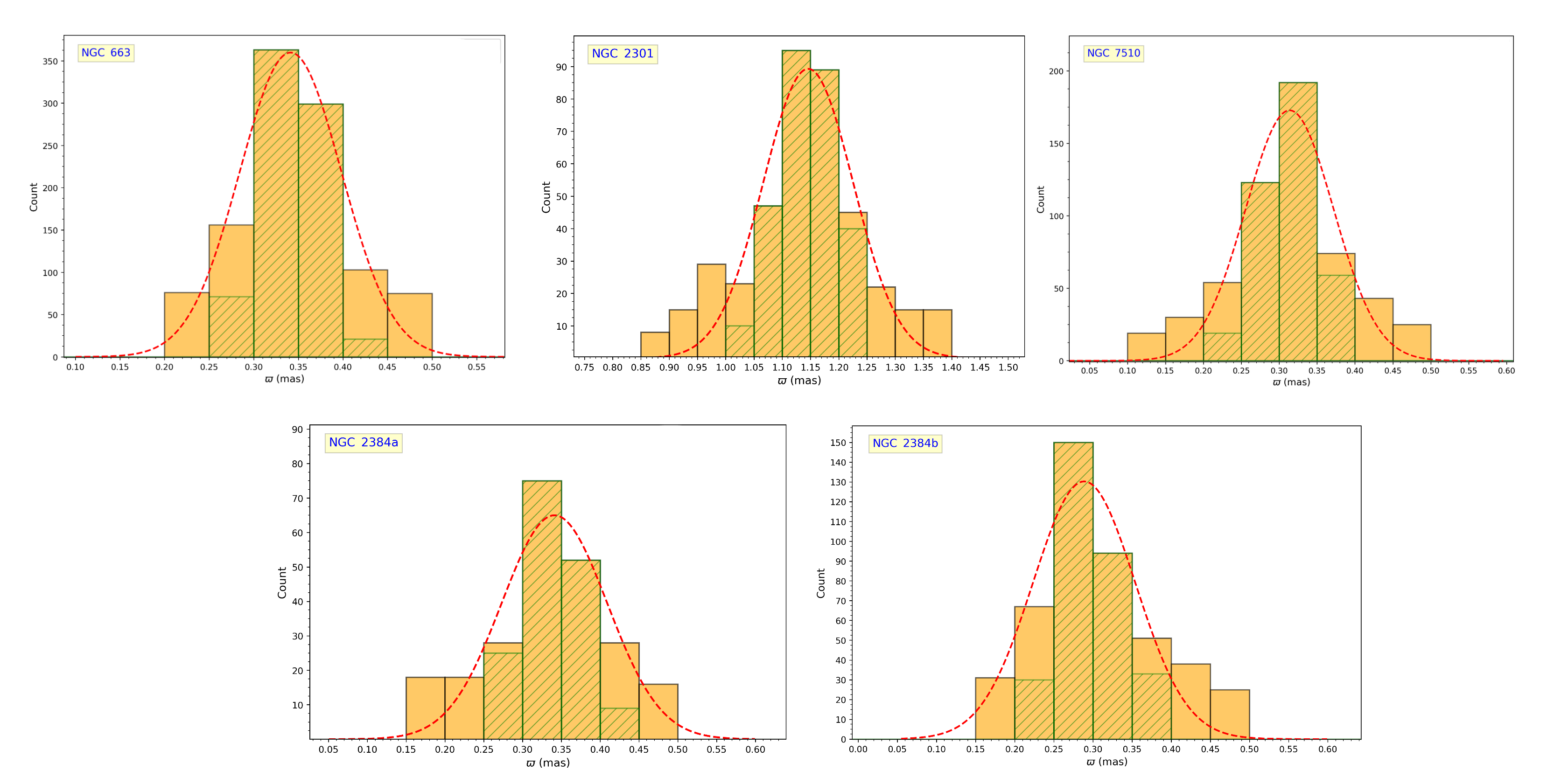} 
\caption{Histograms of the trigonometric parallaxes are shown, with the Gaussian fits overlaid as red dashed lines in each panel. The green shaded regions indicate the 1$\sigma$ parallax intervals.}
\label{fig: Plx}
\end{figure}

We also found that the mean trigonometric parallaxes of probable members are equal to 0.346 $\pm$ 0.0539 (NGC 663), 1.133 $\pm$ 0.1127 (NGC 2301), 0.343 $\pm$ 0.0704 (NGC 2384a), 0.296 $\pm$ 0.0682 (NGC 2384b), and 0.317 $\pm$ 0.0561 (NGC 7510) in mas units, as shown in Figure \ref{fig: Plx}, and their corresponding distances are 3063 $\pm$ 55, 975 $\pm$ 31, 3039 $\pm$ 55, 3477 $\pm$ 59, and 3211 $\pm$ 57 pc with respective manner of NGC 663, NGC 2301, NGC 2384a, NGC 2384b, and NGC 7510 OCs. Our results agree fairly well with the distances reported in the literature for these clusters, as shown in Tables \ref{tab:literature}.

\subsection{Bayesian Inference of Astrophysical Parameters}
\label{sec:astrophysics}

The determination of fundamental parameters such as age, distance, reddening, and metallicity is crucial for understanding the evolutionary state of open clusters. In this study, these astrophysical quantities were derived by comparing the observed \textit{Gaia}-based colour–magnitude diagrams (CMDs) with theoretical stellar models in a fully Bayesian framework. Instead of relying on visual isochrone fitting, we employed a Markov Chain Monte Carlo (\texttt{MCMC}) approach similar to that described in \citet{Tanik2025}, using \texttt{PARSEC} stellar tracks \citep{Bressan_2012} as the underlying evolutionary grid.

The MCMC sampling explores the posterior distribution of the cluster parameters by forward-modelling synthetic magnitudes for a given stellar population and comparing them with the observed photometry of high-probability cluster members. For each trial parameter set, theoretical magnitudes in the $G$, $G_{\rm BP}$, and $G_{\rm RP}$ bands were interpolated from the isochrone grid, and the likelihood was evaluated assuming Gaussian photometric uncertainties. The log-likelihood function has the form:
\begin{equation}
\ln \mathcal{L}(\boldsymbol{\theta}) = 
- \sum_{i=1}^{N_\mathrm{stars}} \sum_{X}
\left[
\frac{(m_{X,i} - \hat{m}_{X,i})^2}{2\sigma_{X,i}^2}
+ \ln(\sqrt{2\pi}\sigma_{X,i})
\right],
\end{equation}
where $\hat{m}_{X,i}$ are model magnitudes and $\boldsymbol{\theta}$ represents the free parameters (age, distance modulus, extinction, and metallicity). The sampling was performed using the \texttt{emcee} package \citep{Foreman-Mackey2013}, with multiple walkers initialized around broad, physically motivated priors. Posterior distributions were obtained after several thousand iterations, and the most probable parameters were adopted together with their 16th–84th percentile uncertainties.

\begin{figure}[h]
\centering
\includegraphics[width=1\linewidth]{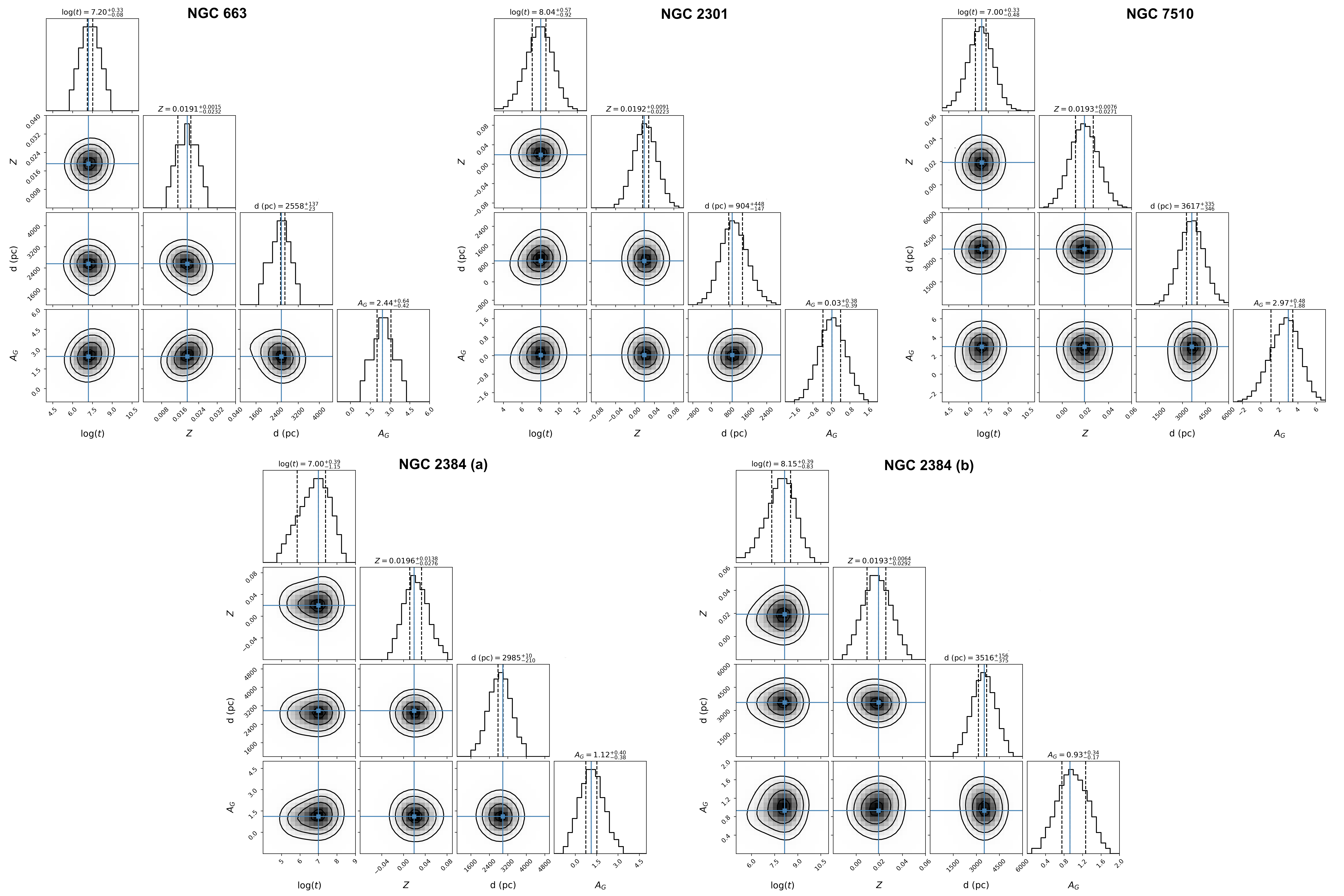} 
\caption{Corner plots display the posterior correlations among the fitted stellar parameters for each cluster following the MCMC analysis. The solid blue markers indicate the optimal values, and the pink dashed lines denote the 16th and 84th percentile limits.}
\label{corner_mcmc}
\end{figure}

\begin{figure}[h]
\centering
\includegraphics[width=1\linewidth]{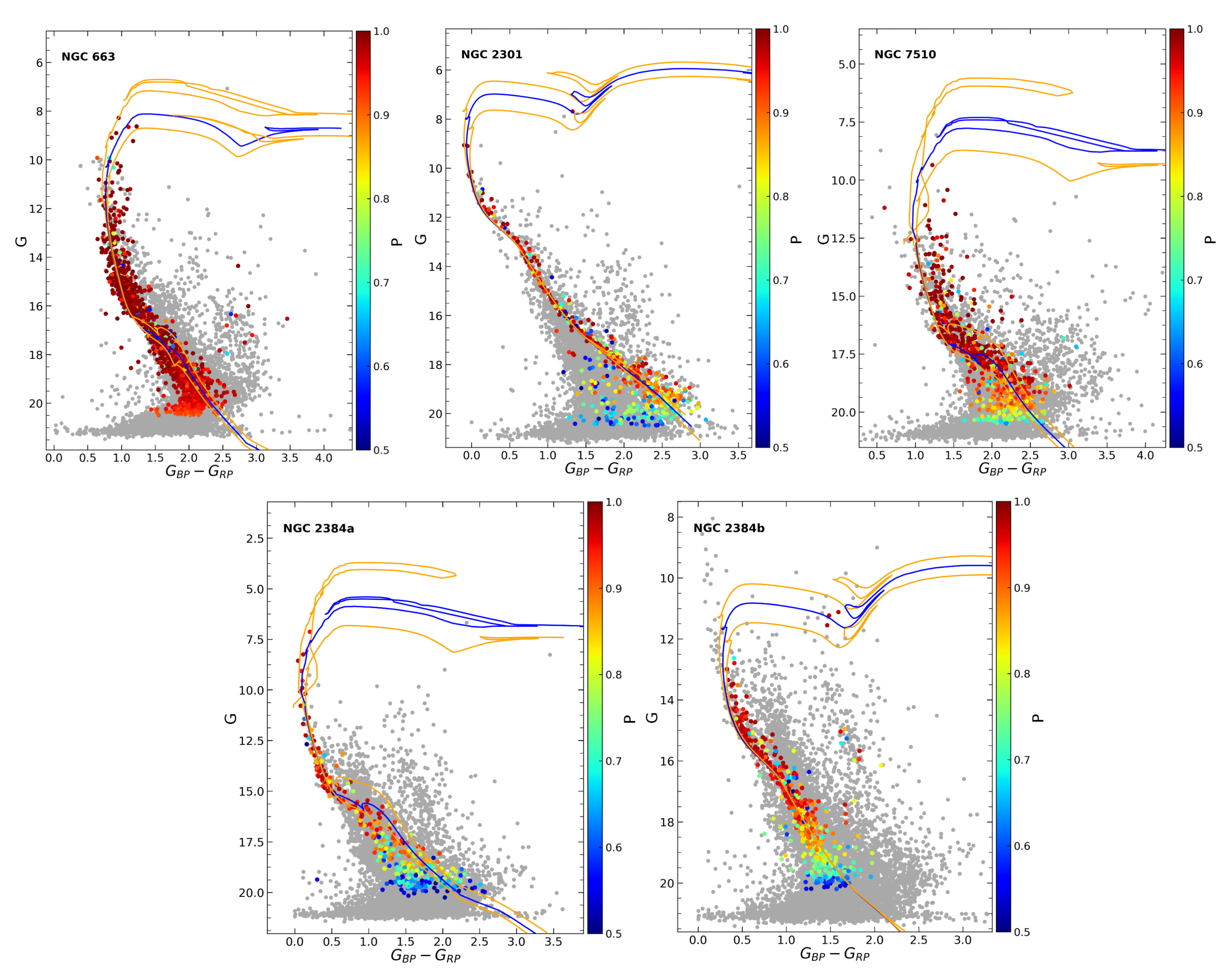} 
\caption{\textit{Gaia} CMDs of the clusters NGC 663, NGC 2301, NGC 7510, NGC 2384a, and NGC 2384b. Gray dots represent field stars, orange dots indicate cluster members, the blue lines show the best-fit isochrones, and the green dashed lines represent the associated uncertainties.
\label{fig: cmd}}
\end{figure}

Figure \ref{corner_mcmc} illustrates the marginal posterior distributions for the five clusters, with parameter uncertainties defined by the 16th and 84th percentile intervals. The observed CMDs of NGC 663, NGC 2301, NGC 2384a, NGC 2384b, and NGC 7510, together with the MCMC-derived best-fitting \texttt{PARSEC} isochrones, are presented in Figure~\ref{fig: cmd}. The posterior distributions obtained from the MCMC sampling provide robust estimates of the cluster ages, reddening values, and distance moduli. Instead of listing these quantities individually in the text, all MCMC-derived parameters for the five clusters, including age, $(m-M)_0$, $E(G_{\rm BP}-G_{\rm RP})$, and the corresponding uncertainties, are summarized in Table~\ref{all_results}. These tabulated values represent the median of the posterior distributions, while the uncertainties correspond to the 16th and 84th percentiles.

As shown in Figure \ref{fig: cmd}, the best-fitting isochrones indicate distinct ages for all clusters from the most likely members (P$\geq$0.5). It can be clearly seen from the CMD that choosing P$\geq$0.5 as a threshold makes sense due to the lower probabilities causing the field contamination. The corresponding distance moduli and colour excesses derived from the fits are not listed individually here, as all values and uncertainties are already reported in Table~\ref{all_results}. The $E(B-V)$ colour excesses were computed using:
\[
E(B-V)=0.775 \times E(G_{BP}-G_{RP}),
\]
and the extinction coefficients using:
\[
A_{\rm G}=2.74 \times E(B-V).
\]
The resulting $E(B-V)$ and $A_G$ values and their uncertainties for all clusters are summarized in Table~\ref{all_results}, along with the photometric distances derived via:
\[
d=10^{((m-M)_{obs}-A_{G}+5)/5}.
\]

The photometric distances, which agree within uncertainties with the astrometric distances, are likewise reported in Table~\ref{all_results}. Our estimated ages for the clusters are generally lower than previously published values (see Table~\ref{tab:literature}), and we also provide, for the first time, an age estimate for the second substructure in NGC 2384. Using the trigonometric parallax distances, we computed the heliocentric Galactic Cartesian coordinates and Galactocentric distances of the clusters using:
\[
X_{\odot} = d \cos{b} \cos{l}, \quad
Y_{\odot} = d \cos{b} \sin{l}, \quad
Z_{\odot} = d \sin{b}.
\]
\[
R_{\rm gc} = \sqrt{R_0^2 + d^2 \cos^2 b - 2 R_0 d \cos b \cos l}.
\]
where $R_0 = 8$ kpc is adopted from \cite{Bovy2015}. All computed positional parameters ($X_\odot$, $Y_\odot$, $Z_\odot$, and $R_{\rm gc}$) are listed in Table~\ref{all_results}.

\begin{table}[t]
\centering
\footnotesize
\caption{Astrophysical and photometric properties of clusters as determined in the present work.
\label{all_results}}
\resizebox{\textwidth}{!}{%
\begin{tabular}{lccccc} 
 \hline
 \textbf{Parameter} & \textbf{NGC 663} & \textbf{NGC 2301} & \multicolumn{2}{c}{\textbf{NGC 2384 (a \& b)}} & \textbf{NGC 7510}  \\
\hline
\hline
Members $(N)$  & 1498 & 575  & 337 & 622 & 723 \\

$\varpi$ (mas) & 0.346 $\pm$ 0.0539 & 1.133 $\pm$ 0.1127   & 0.343 $\pm$ 0.0704  & 0.296 $\pm$ 0.0682 & 0.317 $\pm$ 0.0561 \\ 

d$_\varpi$  (kpc) &  2.890  $\pm$  0.06 & 0.883 $\pm$ 0.02  & 2.916 $\pm$  0.60 & 3.378  $\pm$  0.54 & 3.155  $\pm$  0.56 \\

$\mu _{\alpha}\cos \delta$ (mas yr$^{-1}$)  
& -1.133 $\pm$ 0.0826  
& -1.368 $\pm$ 0.1375 
& -2.298 $\pm$ 0.0885 
& -1.600 $\pm$ 0.0766 
& -3.641 $\pm$ 0.1095  \\  

$\mu_\delta$ (mas yr$^{-1}$) 
& -0.336 $\pm$ 0.0840 
& -2.204 $\pm$ 0.1260 
& 3.121 $\pm$ 0.1033 
& 1.935 $\pm$ 0.0805  
& -2.234 $\pm$ 0.0955\\

$ (m-M)_{obs}$ (mag) 
& 14.80 $\pm$ 0.20 & 10.10 $\pm$ 0.20 & 13.55 $\pm$ 0.20 & 15.40$\pm$0.20 & 15.40 $\pm$ 0.20\\

$E(B-V)$ (mag)
& 0.89$^{+0.07}_{-0.06}$
& 0.09$^{+0.05}_{-0.05}$
& 0.41$^{+0.08}_{-0.09}$
& 0.34$^{+0.09}_{-0.08}$
& 1.08$^{+0.14}_{-0.11}$ \\

$E(G_{BP}-G_{RP})$ (mag)
& 1.15$^{+0.09}_{-0.08}$
& 0.12$^{+0.07}_{-0.06}$
& 0.53$^{+0.10}_{-0.11}$
& 0.44$^{+0.13}_{-0.09}$
& 1.40$^{+0.18}_{-0.14}$ \\

$ A_{G}$ (mag)  
& 2.44$^{+0.18}_{-0.17}$ 
& 0.03$^{+0.14}_{-0.13}$ 
& 1.12$^{+0.22}_{-0.24}$ 
& 0.93$^{+0.27}_{-0.21}$
& 2.97$^{+0.37}_{-0.29}$ \\  

d$_{(m-M)}$ (pc)  
& 2558$^{+137}_{-23}$
& 904$^{+49}_{-17}$  
& 2985$^{+159}_{-136}$ 
& 3516$^{+138}_{-156}$   
& 3617$^{+155}_{-32}$ \\

$Z$ 
& 0.0191$^{+0.0015}_{-0.0023}$ 
& 0.0192$^{+0.0031}_{-0.0020}$ 
& 0.0196$^{+0.0040}_{-0.0038}$ 
& 0.0193$^{+0.0032}_{-0.0028}$ 
& 0.0193$^{+0.0031}_{-0.0021}$ \\

$\log t$ 
& 7.20$^{+0.11}_{-0.13}$
& 8.04$^{+0.18}_{-0.15}$ 
& 7.00$^{+0.12}_{-0.11}$ 
& 8.15$^{+0.17}_{-0.18}$ 
& 7.00$^{+0.13}_{-0.12}$ \\  

$X_{\odot}$ (kpc) & -1.638 $\pm$ 0.20 & -0.759 $\pm$ 0.20 & -1.681 $\pm$ 0.20 & -0.958 $\pm$ 0.20 & -0.958 $\pm$ 0.20  \\ 

$Y_{\odot}$ (kpc) & 1.964 $\pm$ 0.20   & -0.491 $\pm$ 0.20 & -2.465 $\pm$ 0.20 & 2.461 $\pm$ 0.20 & 2.461 $\pm$ 0.20   \\

$Z_{\odot}$ (kpc) & -0.029 $\pm$ 0.20  & 0.013 $\pm$ 0.20 & -0.104 $\pm$ 0.20 & 0.011 $\pm$ 0.02 & 0.011 $\pm$ 0.02 \\

$R_{\rm gc}$(kpc) & 9.836 $\pm$ 0.10 & 8.773 $\pm$ 0.10 & 9.989  $\pm$ 0.10  & 10.393 $\pm$ 0.10 & 9.903 $\pm$ 0.10\\
 \hline    
\end{tabular}}
\end{table}

\section{Luminosity and Mass Functions}
\subsection{Luminosity Function}
\label{sec:LF and MF}

\begin{figure}[htp]
\centering
\includegraphics[width=1\linewidth]{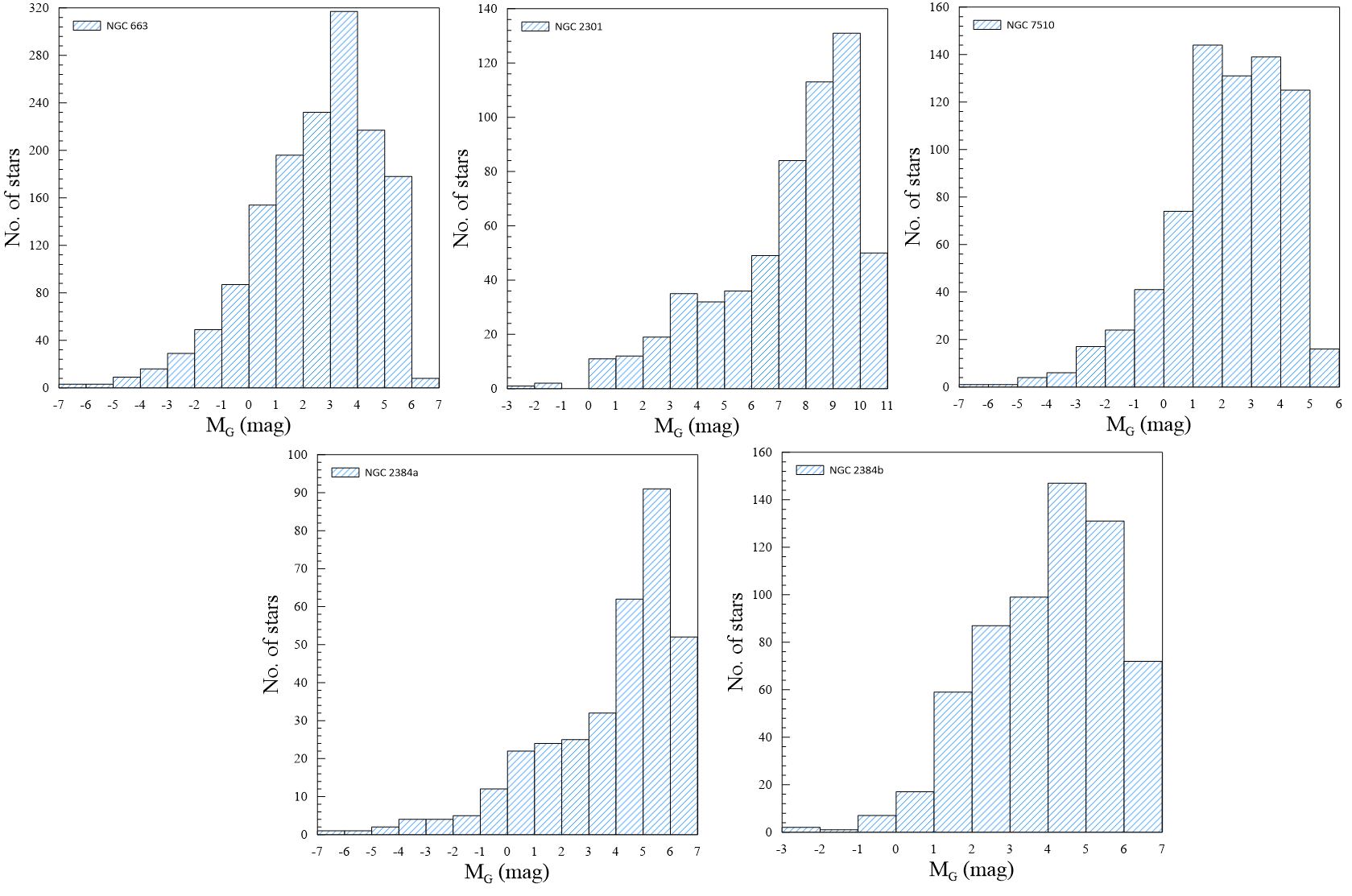}\\
\caption{The LF devoted to NGC 663, NGC 2301, NGC 2384a, NGC 2384b, and NGC 7510.
\label{fig: LFs}}
\end{figure}

The number of stars as a function of brightness (usually expressed in absolute magnitude or apparent magnitude in a certain photometric band) is represented by the luminosity function (LF) of an OC. Particularly when distances and reddening are well-constrained, the LF acts as an observable proxy for the mass function and is intimately related to it. To create accurate LFs, modern studies make use of extensive astrometric and photometric surveys like $Gaia$ DR3 \citep[e.g.][]{GaiaDR3}. With the use of proper-motion components, trigonometric parallax, and $V_{\rm r}$, these datasets enable accurate membership determination. For instance, \citet{Cantat_2020} improved our knowledge of cluster evolution and Galactic structure by using $Gaia$ data to generate high-precision LFs for hundreds of clusters. Figure \ref{fig: LFs} shows the LF which reflects the incompleteness in the photometry in the ranges of $M_{\rm G}$ from -7.00 to 7.00 mag (NGC 663), from -3.00 to 11.00 mag (NGC 2301), from -7.00 to 7.00 mag (NGC 2384a), from -3.00 to 7.00 mag (NGC 2384b), and from -7.00 to 6.00 mag (NGC 7510).

\subsection{Mass Function}

\begin{figure}[htp]
\centering
\includegraphics[width=1\linewidth]{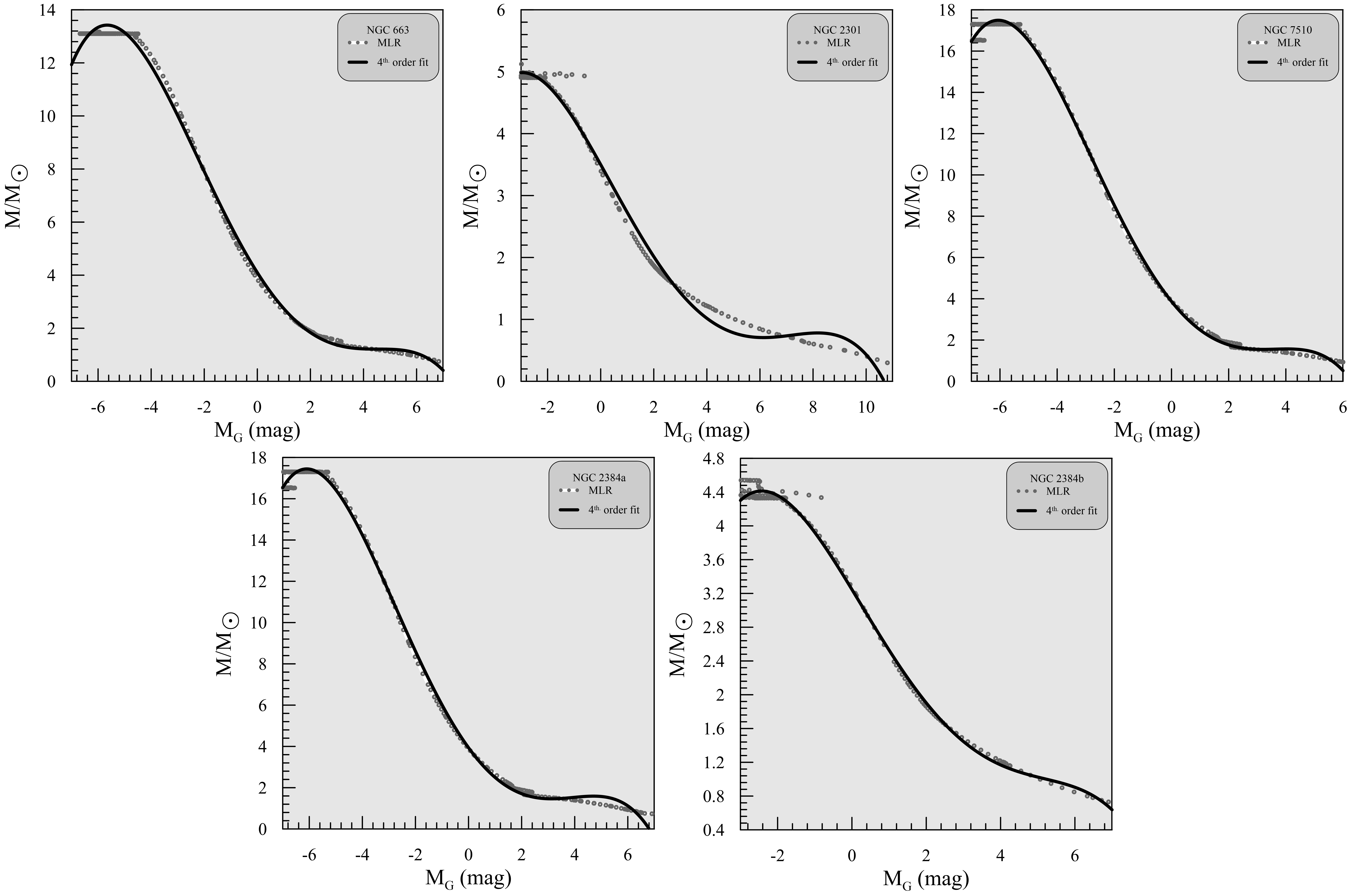}\\
\caption{Mass–absolute magnitude ($M/M_{\odot}$–$M_G$) relations from PARSEC isochrones, computed using the metallicities derived for the clusters in this study.
\label{fig: MLR}}
\end{figure}

The Initial Mass Function (IMF), which is a direct result of stellar evolution, acts as essential data on the distribution of stellar masses \citep{Bisht2019, 2025arXiv251203552B}. \citet{Salpeter1955} established the IMF for stars in the solar vicinity and suggested a power-law model for the MF as follows:
\begin{equation}
\dfrac{dN}{dM} \propto M^{-\alpha},
\label{sec:MF}
\end{equation}
$dN/dM$ here refers to the number of stars in the mass range $(M, M + dM)$ and $\alpha$ is a constant parameter, conventionally set to 2.35. This value, referred to as the Salpeter slope, signifies that low-mass stars are considerably more abundant than their high-mass counterparts. 
In the present work, the absolute magnitudes and the masses of the adopted isochrones \citep{2018A&A...616A...4E} with metallicities; $Z = 0.0191^{+0.0015}_{-0.0023}$ for NGC 663, $Z = 0.0192^{+0.0031}_{-0.0020}$ for NGC 2301, $Z = 0.0196^{+0.0040}_{-0.0038}$ for NGC 2384a, $Z = 0.0193^{+0.0032}_{-0.0028}$ for NGC 2384b, and $Z = 0.0193^{+0.0031}_{-0.0021}$ for NGC 7510 OCs are used to construct a relation between $M/M_{\odot}$ and absolute magnitude $M_G$ and fitting like the fourth order, and the resulting results are shown in Figure \ref{fig: MLR} with numerical values listed in Table \ref{MFs}.

The stellar masses within the clusters are derived by converting LFs into MFs through the mass-luminosity relation (MLR) based on the selected isochrones with \cite{Elsanhoury2018} and fitting like the fourth order, and the resulting results are shown in Figure \ref{fig: MLR}.

i.e., 
\begin{equation}
M_{\rm c} = a_0 + a_1 \times M_{\rm G} + a_2 \times M_{\rm G}^2 + a_3 \times M_{\rm G}^3 + a_4 \times M_{\rm G}^4,
\label{Eq: MLR}
\end{equation}
where the coefficients $a_0,~a_1,~a_2,~a_3$, and $a_4$ are those that result from fitting the isochrones to each cluster's MF. These characteristics are essential for precisely estimating star masses and provide important information about the clusters' stellar populations. Eq. \ref{Eq: MLR} is used to get the estimated total mass ($M_{\text{C}}$) of the clusters NGC 663, NGC 2301, NGC 2384a, NGC 2384b, and NGC 7510, which are approximately 3584 $\pm$ 60, 486 $\pm$ 22, 725 $\pm$ 27, 832 $\pm$ 29, and 1863 $\pm$ 43 $M_{\odot}$, respectively, as a function of the previously determined ages and metallicity.

\begin{figure}[htp]
\centering
\includegraphics[width=1\linewidth]{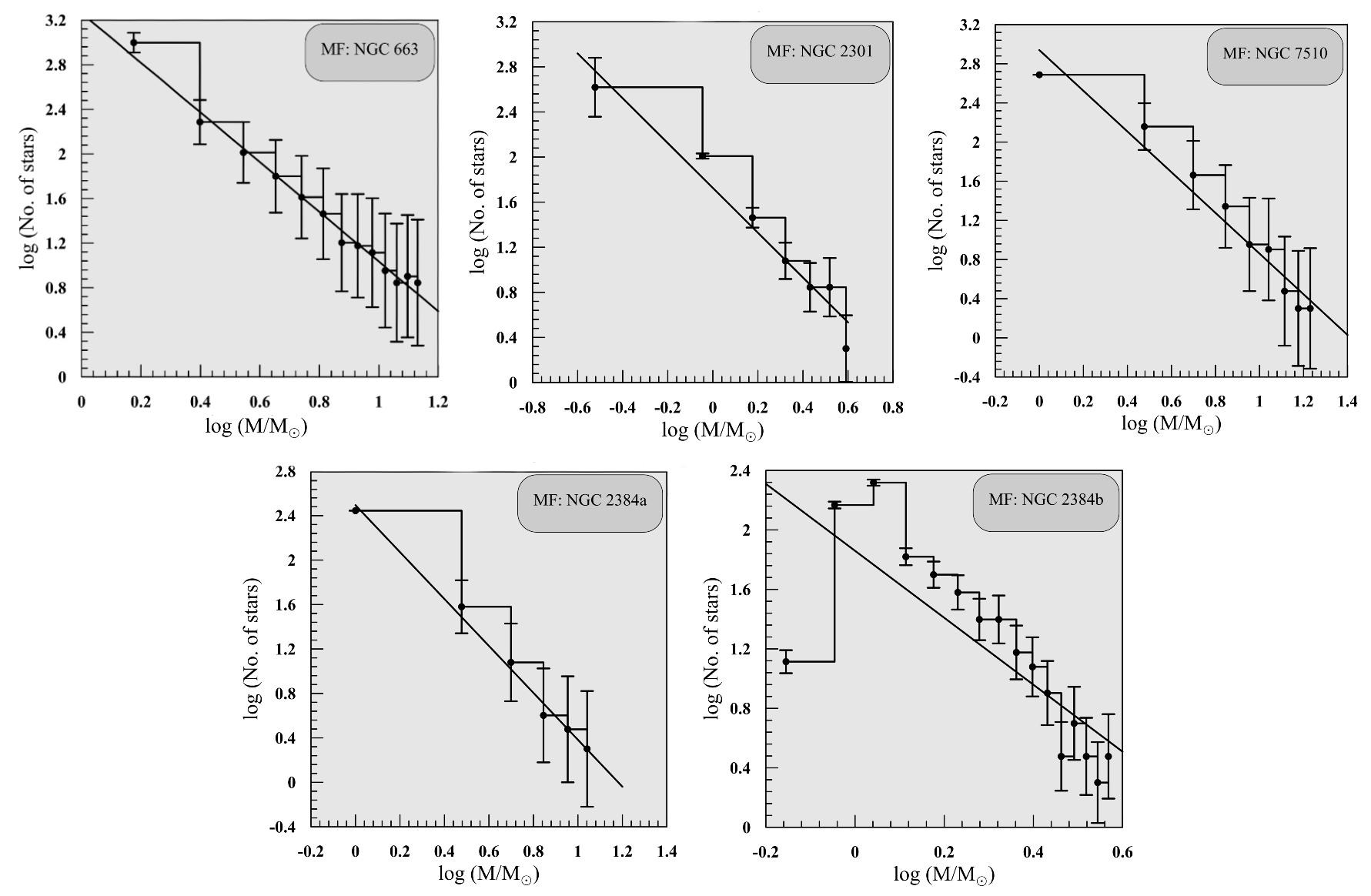}\\
\caption{The cluster's MFs fitted with the Salpeter power-law line reveal the characteristics of the mass distribution.
\label{fig: MFs}}
\end{figure}

The key parameters characterizing the stellar populations and physical properties of the OCs are summarized in Table \ref{MFs}, which includes the coefficients $a_0, a_1, a_2, a_3$, and $a_4$, $M_{\text{C}}$, and mean mass $\langle M_C \rangle$. These values provide valuable insight into the structure and evolutionary state of the clusters. Significant variations are observed among the clusters, not only in terms of their total mass but also in the number of stars and the mean stellar mass per OC. These differences reflect the heterogeneity in their stellar populations and the various stages of evolution they may be undergoing. $\langle M_C \rangle$ of the stars in each OC offers a further understanding of the distribution of stellar masses within the population. To visualize the mass distribution across the clusters, we have plotted the MFs of these OCs on a logarithmic scale, as shown in Figure \ref{fig: MFs}. The MF slopes ($\alpha$) of the linear fits for each OC can be derived by using Eq. \ref{sec:MF} and are listed in Table \ref{MFs}, which reflects good agreement with \citet{Salpeter1955} value.

As shown with many authors \citep[e.g.,][]{2003PASP..115..763C, 2001MNRAS.322..231K, 1986FCPh...11....1S, 1979ApJS...41..513M}, The available constraints can be conveniently summarized by the multiple-part power-law IMF is:
\begin{equation}
    \xi(m) \propto m^{-\alpha_i}=m^{\gamma_i}
\end{equation}
where
\begin{equation}
\label{Kroupa}
\begin{split}
\alpha_o & = +0.3~\pm~0.7,~0.01 \le m/M_{\odot} <~0.08,  \\
&\alpha_1 = +1.3~\pm~0.5,~0.08 \le m/M_{\odot} <~0.50, \\
&\alpha_2  = +2.3~\pm~0.3,~0.50 \le m/M_{\odot} <~1.00, \\
&\alpha_3  = +2.3~\pm~0.7,~1.00 \le m/M_{\odot}. \\
\end{split}
\end{equation}
and $\xi(m)~dm$ is the number of single stars in the mass interval $m$ to
$m \pm dm$. Therefore, its clearly that for our program clusters its consistent with $\alpha_2$ and $\alpha_3$ with Equations \ref{Kroupa}.

\begin{table}[t]%
\centering
\caption{The MFs and their derived parameters.}
\begin{tabular}{l|ccccc}
\hline
\textbf{Parameters} & \textbf{NGC 663}  & \textbf{NGC 2301}  & \textbf{NGC 2384a} & \textbf{NGC 2384b} & \textbf{NGC 7510} \\
\hline
\hline
$M_{\text{C}}$ ($M_{\odot}$) & 3584 $\pm$ 60 & 486 $\pm$ 22  & 725 $\pm$ 27 & 832 $\pm$ 29 & 1863 $\pm$ 43   \\
$\langle M_C \rangle$ ($M_{\odot}$) & 2.40   & 0.85  & 2.15 & 1.34 & 2.58  \\
$\alpha$ & 2.23 $\pm$ 0.070  &  2.00 $\pm$ 0.030 & 2.12 $\pm$ 0.070 & 2.26 $\pm$ 0.070 & 2.08 $\pm$ 0.070  \\
$a_0$ &  4.098 $\pm$ 0.002  &  3.500 $\pm$ 0.005  & 3.937 $\pm$ 0.003 & 3.247 $\pm$ 0.002 & 3.871 $\pm$ 0.002 \\
$a_1$ & -1.580 $\pm$ 0.002  & -0.770 $\pm$ 0.002  & -1.772 $\pm$ 0.002 & -0.725 $\pm$ 0.002 & -1.725 $\pm$ 0.002\\
$a_2$ &  0.208 $\pm$ 0.003  & -0.021 $\pm$ 0.003  & 0.326 $\pm$ 0.001 & -0.018 $\pm$ 0.001 & 0.341 $\pm$ 0.001 \\
$a_3$ &  0.016 $\pm$ 0.002  &  0.020 $\pm$ 0.002  & 0.012 $\pm$ 0.004 & 0.028 $\pm$ 0.001 & 0.009 $\pm$ 0.001\\
$a_4$ & -0.004 $\pm$ 0.001  & -0.002 $\pm$ 0.001  & -0.005 $\pm$ 0.001 & -0.003 $\pm$ 0.001 & -0.006 $\pm$ 0.001\\
\hline
\end{tabular}
\label{MFs}
\end{table}

\section{Dynamical and kinematical structure}
\label{sec:kinematics}

According to recent research, the majority of clusters have roughly Gaussian star velocity distributions with tiny internal dispersions (0.1–0.5 km s$^{-1}$), which are typical of virial equilibrium systems \citep{Soubiran_2018}. But in several young clusters, substructures like streaming movements and kinematic clumps have been observed, indicating that many clusters originate with complex velocity structures \citep{KounkelCovey2019}. Some OCs have also been observed to rotate. According to \citet{2021A&A...654A.122G}, the Hyades and Praesepe exhibit indications of light internal rotation, which may be a result of tidal interactions or a reminder from their formation process.

The gravitational interactions between member stars influence the internal dynamics of OCs. Star velocities redistribute toward equilibrium as a result of dynamical relaxation caused by these interactions throughout time. The size, mass distribution, and number of stars in a cluster all affect the relaxation period, or the timescale over which it achieves energy equipartition. By time ranging from tens to hundreds of Myr, the majority of OCs relax \citep{binney08}. In addition to interacting with spiral arms, molecular clouds, and the Galactic tidal field, OCs are not solitary systems. Cluster stars may be disturbed by these outside pressures, which could result in a slow loss of mass. It is thought that encounters with massive molecular clouds are the primary cause of cluster breakdown and are especially disruptive \citep{Spitzer1971, 2020MNRAS.498.2472K}.

\subsection{The Dynamical Relaxation Time}
\label{sec:dynamic and kinematic}
The period of time over which a star in a cluster undergoes sufficiently weak gravitational interactions with other stars to cause a significant change in its velocity vector is known as the relaxation time ($T_{\rm relax}$). During this period, the cluster evolves from a beginning, potentially asymmetrical condition to a dynamically relaxed configuration. Following the prescription of \citet{Spitzer1971}, $T_{relax}$ is calculated using the relation:
\begin{equation}
\label{Eq: t_relax}
T_\text{relax} = \frac{8.9 \times 10^5 N^{1/2} R_\text{h}^{3/2}}{\langle M_C \rangle^{1/2} \log(0.4N)},
\end{equation}
where $N$ represents the cluster's total number of stars, $R_{\rm h}$ is the half-mass radius (pc), and $\langle M_C \rangle$ represents the mean cluster mass. \citet{Sableviciute_2006} given a formula that was used to derive the half-mass radius:
\begin{equation}
R_{\rm h} = 0.547 \times r_{\rm c} \times \left( \frac{r_{\rm cl}}{r_{\rm c}} \right)^{0.486},
\end{equation}
where $r_{\rm cl}$ (pc) is the cluster radius derived from the RDP analysis, as reported in Table~\ref{table-king}. Therefore, the obtained results are 9.19 $\pm$ 3.03 (NGC 663), 3.55 $\pm$ 1.88 (NGC 2301), 12.89 $\pm$ 3.60 (NGC 2384a), 12.89 $\pm$ 3.60 (NGC 2384b), and 10.33 $\pm$ 3.21 (NGC 7510). Using this approach. On the other hand, the calculated $R_{\rm h}$ (pc) and $T_{\rm relax}$ are listed here with Table \ref{tab:full_parameters}.

The dynamical evolution parameter, $\tau = \text{age} / T_{\text{relax}} $, may be utilized to characterize the dynamical state of clusters \citep{Bisht2021}. This parameter is significantly greater than unity ($ \tau \gg 1 $) for relaxed clusters and vice versa for non-relaxed ones. The period during which all member stars are expelled as a result of internal stellar interactions is represented by the evaporation time, which is $ \tau_{\text{ev}} \simeq 10^2 T_{\text{relax}}$ (in Myr) \citep{2001ApJ...553..744A}. Therefore, low-mass stars usually escape through the Lagrange points in this mechanism at low velocities \citep{koposov2008}. Our obtained results are summarized in Table \ref{tab:full_parameters}, which reflects that these clusters are non-relaxed ones except NGC 2301 and NGC 2384b.

\subsection{The Convergent Point}
A point on the celestial sphere known as the Convergent Point (CP) \citep{Elsanhoury2018, Elsanhoury2025, Elsanhoury2025b, Bisht2022a, Bisht2022b} is where, because of perspective, the PM vectors of stars in a co-moving group—like an open cluster—appear to converge. The CP is the point at which the observed PMs of a cluster of stars appear to converge when projected onto the sky if they are travelling almost parallel through space. In terms of mathematics, this idea can be linked to spherical trigonometry and vector motion equations \citep{2012A&A...538A..23G}.

\begin{figure}[ht]
\centering
\includegraphics[width=0.99\linewidth]{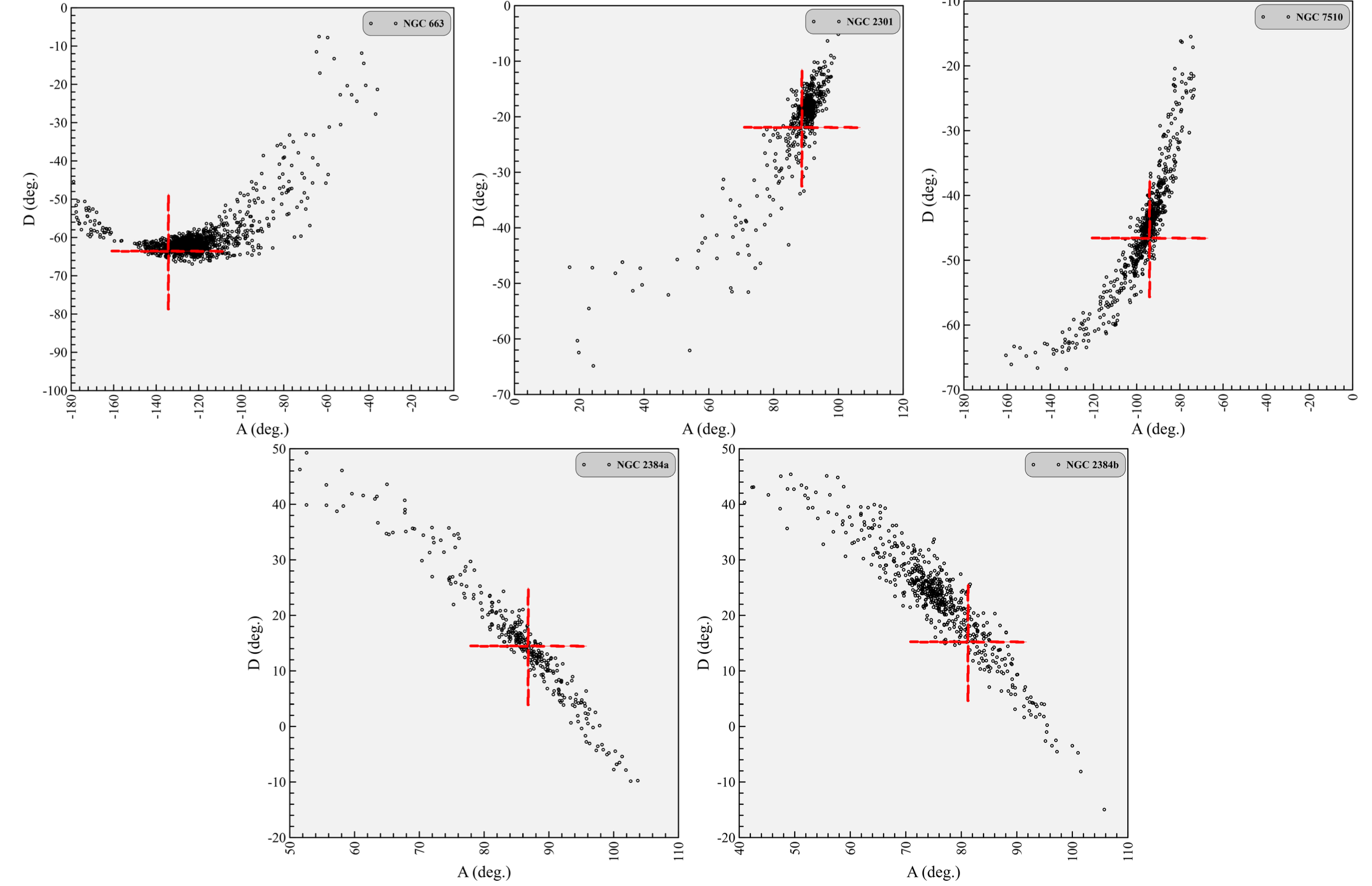}

\caption{AD-diagram for NGC 663, NGC 2301, NGC 7510, NGC 2384a, and NGC 2384b, where the cross symbols denote the positions of the calculated apex coordinates $(A_0,~D_0)$.}
\label{fig: CP}
\end{figure}

Finding the apex provides important insights into the kinematic structure and overall dynamical coherence of the cluster. The AD-diagram approach, which was originally laid out by \citet{Chupina2001, Chupina2006}, is frequently employed to do this. Using this method, the apex coordinates ($A_o$, $D_o$) that are determined from the velocity vectors of individual stars are shown in Figure \ref{fig: CP}. The right ascension and declination of the convergent point are indicated by $A_o$ and $D_o$, respectively.

\begin{equation} \label{Eq: 7}
A_o = \tan^{-1}\left(\dfrac{\overline{V}_y}{\overline{V}_x}\right),
\end{equation}
\begin{equation} \label{Eq: 8}
D_o = \tan^{-1} \left(\dfrac{\overline{V}_z}{\sqrt{\overline{V}^{2}_{x} + \overline{V}^{2}_{y}}}\right),
\end{equation}
where $\overline{V}_x$, $\overline{V}_y$, and $\overline{V}_z$ are the space velocity components along the respective axes. Using the formula given by \citet{Melchior1958}, the space velocity components $V_x$, $V_y$, and $V_z$ are then computed.

\begin{equation} \label{Eq. 9-10-11}
\begin{pmatrix}
V_x \\\\
V_y \\\\
V_z
\end{pmatrix} 
= 
\begin{pmatrix}
-4.74~d~\mu_\alpha\cos{\delta}\sin{\alpha} - 4.74~d~\mu_\delta\sin{\delta}\cos{\alpha} \\
+ V_{\rm r} \cos\delta \cos\alpha \\
+4.74~d~\mu_\alpha\cos{\delta}\sin{\alpha} - 4.74~d~\mu_\delta\sin{\delta}\cos{\alpha} \\
+ V_{\rm r} \cos\delta \cos\alpha \\
+4.74~d~\mu_\delta\cos\delta + V_{\rm r}\sin\delta
\end{pmatrix}
\end{equation}

Radial velocity $(V_r)$ data are essential for performing detailed kinematic and orbital dynamical analyses of OCs. In this study, the mean radial velocities $\langle V_r \rangle$ of the clusters NGC 663, NGC 2301, NGC 2384a, NGC 2384b, and NGC 7510 were estimated using $Gaia$ DR3 measurements. To ensure a reliable membership selection, only stars with a cluster membership probability $P~\ge~0.5$ were included in the analysis. $Gaia$ DR3 provided $V_{\rm r}$ measurements for the clusters NGC 663, NGC 2301, NGC 2384a, NGC 2384b, and NGC 7510 from 3, 94, 3, 5, and 2 stars, respectively.

$\langle V_r \rangle$ for each cluster was computed using a weighted mean approach, following the methodology outlined by \cite{Soubiran_2018}. The resulting values yielded $V_r=-65.00\pm0.87$ km s$^{-1}$ for NGC 663, $V_r=25.89\pm0.12$ km s$^{-1}$ for NGC 2301, $V_r=55.80\pm3.76$ km s$^{-1}$ for NGC 2384a, $V_r=71.58\pm0.17$ km s$^{-1}$ for NGC 2384b, and $V_r=-43.81\pm5.08$ km s$^{-1}$ for NGC 7510. Therefore, the obtained results are seen in Table \ref{tab:full_parameters} and the identification of the apex positions of the CP coordinates are seen in Figure \ref{fig: CP} for the respective manner of these corresponding OCs, which were determined using the AD-diagram method.

The transformation equations are defined in Eqs. (\ref{Eq. 12}) were used to translate the space velocity components from the equatorial coordinate system to the Galactic frame. The velocity components ($V_x$, $V_y$, $V_z$) can be translated into their Galactic counterparts, $U$, $V$, and $W$, which represent the velocity vector components with respect to the Galactic centre, using these equations. These components were calculated using the following expressions:
\begin{equation} \label{Eq. 12}
\begin{pmatrix}
U \\\\
V \\\\
W
\end{pmatrix} 
= 
\begin{pmatrix}
-0.0518807421 \; V_{x} -
0.872222642 \; V_{y} -\\
0.4863497200 \; V_{z} \\
+0.4846922369 \; V_{x} -
0.4477920852 \; V_{y} +\\
0.7513692061 \; V_{z}\\
-0.873144899 \; V_{x} -
0.196748341 \; V_{y} +\\
0.4459913295 \; V_{z} \\
\end{pmatrix}
\end{equation}

\begin{figure*}[ht]
\centering
\includegraphics[width=0.75\linewidth]{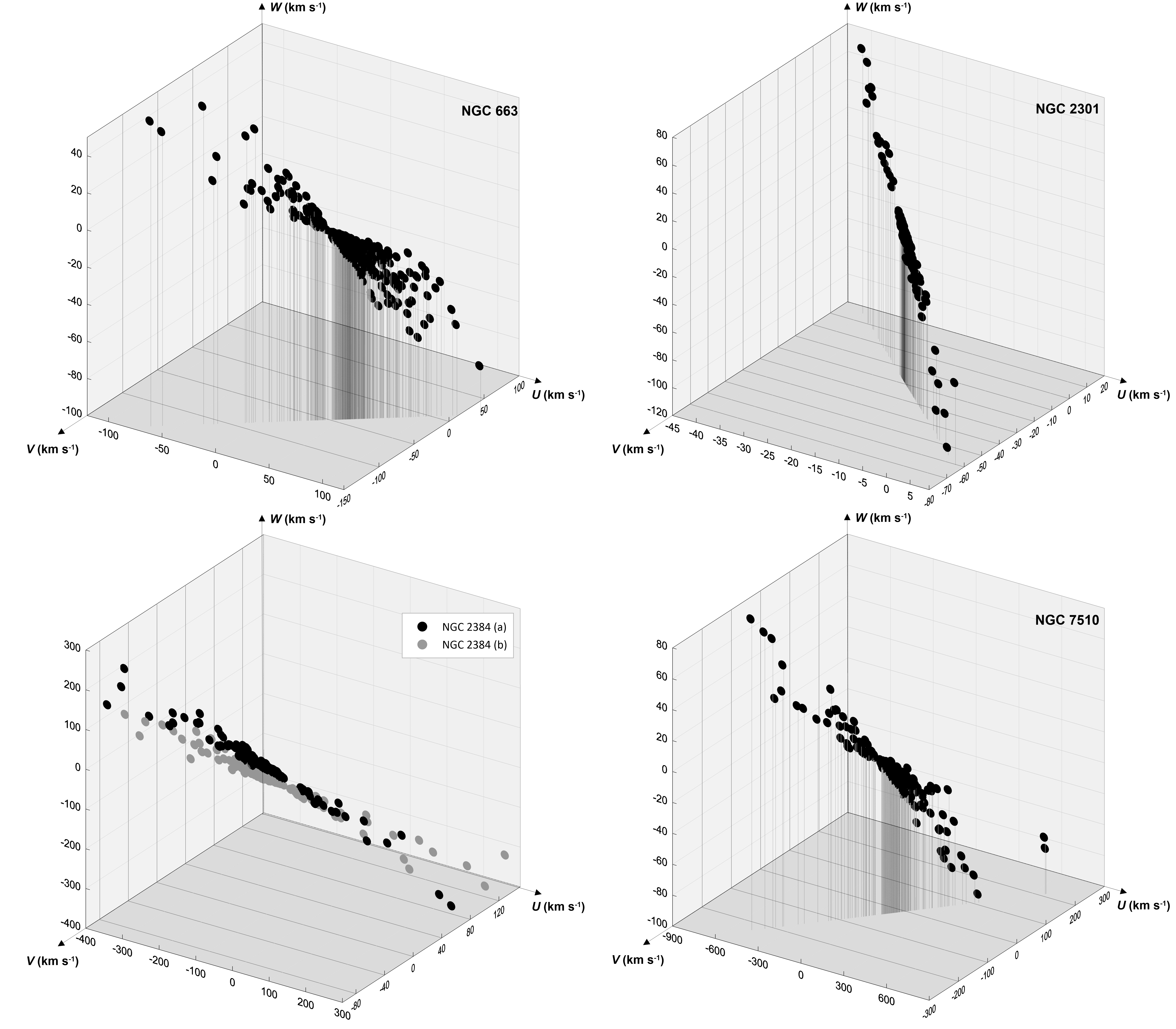}
\caption{3D spatial velocity component distribution for the clusters under investigation projected onto the Galactic coordinate system.}
\label{fig: UVW}
\end{figure*}

By averaging the individual velocity components of each star in the cluster, the mean Galactic space velocity components $\overline{U}$, $\overline{V}$, and $\overline{W}$ are thus determined. To express these average velocities, the following formulas are used:
\begin{equation} \label{eq: 15}
\overline{U} = \dfrac{1}{N}\sum ^{N}_{i=1}U_i,\;\;
\overline{V} = \dfrac{1}{N}\sum ^{N}_{i=1}V_i, \;\; \text{and} \; \; \overline{W} = \dfrac{1}{N}\sum ^{N}_{i=1}W_i,
\end{equation}
where $N$ represents the total number of stars in the cluster. The spatial velocity component distribution for each Galactic member star within the clusters is shown in Figure \ref{fig: UVW}. Additionally, Table \ref{tab:full_parameters} displays the average space velocity components $(U,~V,~W)$ for the clusters. 

\subsubsection{Solar Motion Elements}

The Sun's velocity in relation to the Local Standard of Rest (LSR), a reference frame that shows the standard motion of nearby stars, is referred to as solar motion. Due to their uniform kinematics and extensive distribution throughout the Galactic disk, OCs, which are made up of stars that have a shared origin, age, and velocity, are useful tracers for figuring out solar motion. The mean spatial velocity components ($\overline{U},~\overline{V},~\overline{W}$) of a particular stellar cluster in Galactic coordinates can be used to calculate the solar space velocity components, expressed in km s$^{-1}$. The following relations are used to determine these components \citep{Elsanhoury2016, Elsanhoury2022, Haroon2025}:
\begin{equation} \label{Eq.16}
U_{\odot} = -\overline{U}, \; V_{\odot} = -\overline{V}, \; \text{and} \; W_{\odot} = -\overline{W}.
\end{equation}
The following expression is then used to calculate the solar space velocity's magnitude in relation to the observed objects:
\begin{equation} \label{Eq. 17}
S_{\odot}=\sqrt{(\overline{U})^2+(\overline{V})^2+(\overline{W})^2}.
\end{equation}

Through this calculation, the solar apex in Galactic coordinates can be found using the following formulas:
\begin{equation} \label{Eq. 18}
l_\text{A} = \tan^{-1}\left(\frac{-\overline{V}}{\overline{U}}\right) \;\; \text{and} \;\;
b_\text{A} = \sin^{-1} \left(\frac{-\overline{W}}{S_\odot}\right),
\end{equation}
Here, $l_\text{A}$ and $b_\text{A}$ mark the solar apex's latitude and Galactic longitude, respectively.  For every cluster under analysis, we determined the solar space velocity and associated apex coordinates using these formulas. Table \ref{tab:full_parameters} provides a summary of the findings.

\subsection{Dynamical Orbit Analyses}
\label{sec:orbit}

The Galactic orbital analyses were carried out using both axisymmetric and non-axisymmetric gravitational models implemented in the {\sc galpy}\footnote{\url{https://galpy.readthedocs.io/en/latest/}} package. The widely used axisymmetric model {\sc MWPotential2014} \citep{Bovy2015} was adopted as the baseline potential. This model provides an observationally calibrated description of the Milky Way’s gravitational field and consists of three components: a spherical bulge with a power-law profile and exponential cut-off, a flattened disc following the Miyamoto Nagai formulation, and a spherical dark matter halo represented by a Navarro Frenk White (NFW) profile. The total axisymmetric potential can be written as:
\begin{equation}
\Phi_{\rm MW}(R,z) = \Phi_{\rm bulge}(r) + \Phi_{\rm disc}(R,z) + \Phi_{\rm halo}(r),
\end{equation}
where $r=\sqrt{R^2+z^2}$ is the Galactocentric spherical radius. This formulation reproduces key Galactic observables, such as the rotation curve and the local circular velocity, making it well-suited for orbit integrations.

However, an axisymmetric potential alone does not capture the effects of asymmetric structures such as spiral arms, which play a significant role in the long-term orbital evolution of clusters. Since the studied clusters NGC\,663 NGC\,2301 NGC\,2384, and NGC\,7510 are located in or near major spiral arm segments (Fig. \ref{fig: DSS}), we extended the analysis by incorporating a non-axisymmetric contribution using the {\sc SpiralArmsPotential} \citep{Bovy2015}. This potential introduces a logarithmic spiral perturbation characterized by the number of arms $m$, pitch angle $i$, pattern speed $\Omega_{\rm sp}$, and amplitude $A_{\rm sp}$. The spiral perturbation is expressed as:
\begin{equation}
\Phi_{\rm sp}(R,\phi,z,t) = A_{\rm sp} \cos \Bigg[ m \Big( \phi - \frac{\ln(R/R_0)}{\tan i} - \Omega_{\rm sp} t \Big) \Bigg] , f(z),
\end{equation}
where $R_0$ is a reference radius and $f(z)$ accounts for the vertical attenuation. The total potential in the non-axisymmetric case becomes:
\begin{equation}
\Phi_{\rm total}(R,\phi,z,t) = \Phi_{\rm MW}(R,z) + \Phi_{\rm sp}(R,\phi,z,t).
\end{equation}

As input for the orbital integrations, we used the equatorial coordinates $(\alpha,\delta)$, mean proper motions ($\langle \mu_{\alpha}\cos\delta \rangle$, $\langle \mu_\delta \rangle$), heliocentric distances, radial velocities $V_{\rm r}$, and their uncertainties. A Galactocentric distance of $R_{\rm gc}=8$ kpc and a circular velocity of $V_{\rm rot}=220$ km s$^{-1}$ were adopted for the Sun \citep{Bovy2015,Bovy2012}. Orbits were integrated backward in time for 3 Gyr with a time resolution of 1.5 Myr, in order to trace the clusters’ past dynamical histories \citep{Cinar_2025, Canbay_2025, Alzahrani2025a, Tasdemir2025b}. From these calculations, we obtained the orbital parameters $R_a$, $R_p$, $e$, $Z_{\rm max}$, $(U_{\rm LSR},V_{\rm LSR},W_{\rm LSR})$, and $T_p$, which are listed in Table~\ref{tab:full_parameters}.

\begin{figure}
\centering
\includegraphics[width=0.95\linewidth]{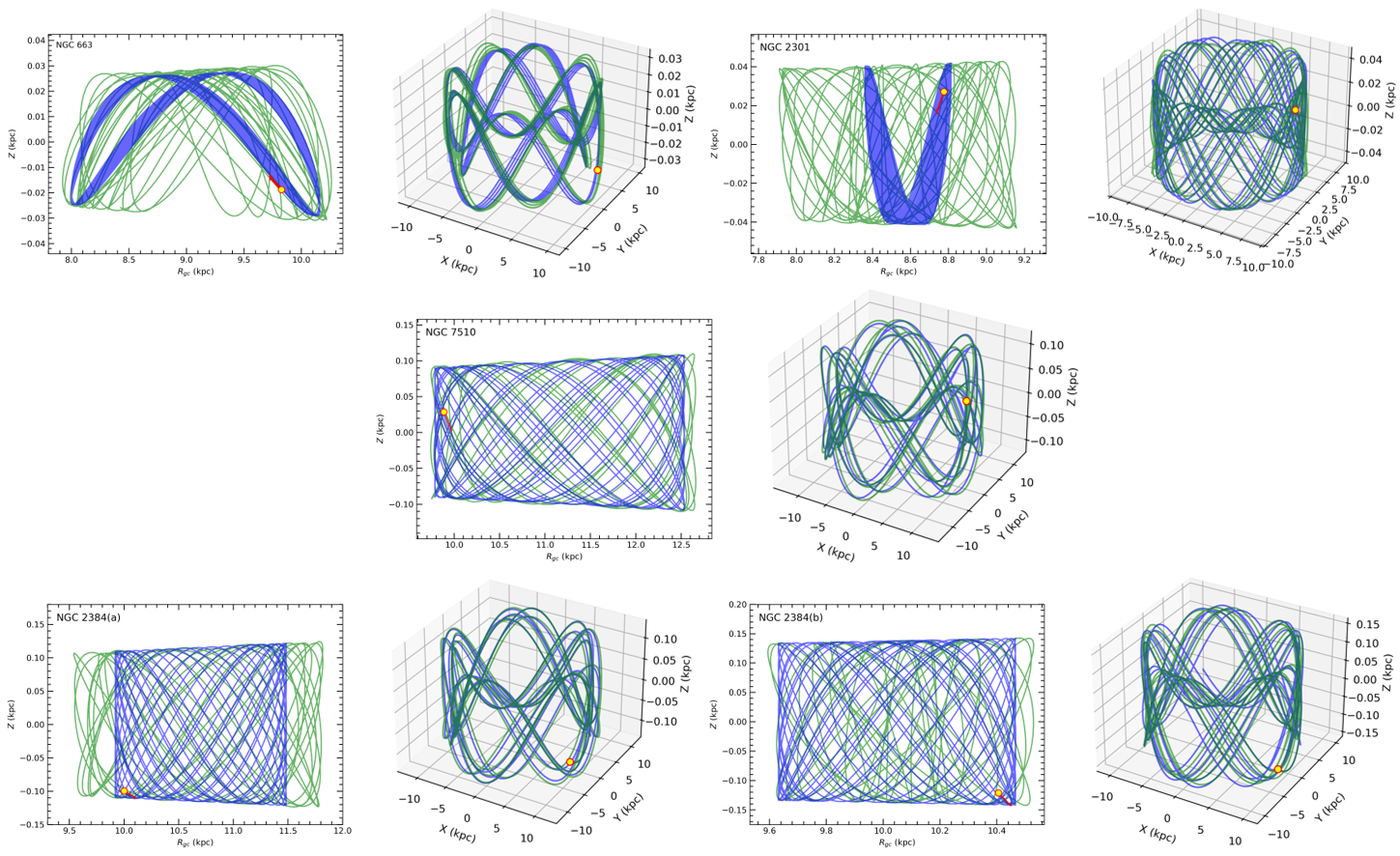}
\caption{The Galactic orbits and radial evolution of the open clusters NGC 663, NGC 2301, NGC 2384a, NGC 2384b, and NGC 7510 are shown in the $Z\times R_{\rm gc}$ (left),  and 3D spatial (right) planes. The yellow-filled circles represent the present-day positions of the clusters, while the yellow-filled triangles indicate their estimated birth positions. The red arrows show the current motion vectors. Orbital trajectories were computed under both the axisymmetric {\sc MWPotential2014} and the MW + Spiral potentials ({\sc SpiralArmsPotential}, \citet{Bovy2015}). In each panel, the trajectory computed with {\sc MWPotential2014} is shown in blue, and the trajectory with MW + Spiral is shown in green, illustrating the differences in orbital evolution between the two models.}
\label{fig: orbital}
\end{figure}

Figure~\ref{fig: orbital} presents the resulting orbital trajectories: the $Z \times R_{\rm gc}$ plane shows a side view of the motion relative to the Galactic plane. The impact of observational uncertainties on the orbital solutions is also highlighted in the Figure \ref{fig: orbital}.

The comparison shows that orbits integrated in the {\sc MWPotential2014} model are generally more regular and symmetric, whereas the inclusion of spiral arms introduces perturbations that lead to more complex trajectories.To test how sensitive our results are to the adopted Galactic potential, we integrated the orbits using both the axisymmetric {\sc MWPotential2014} and a non-axisymmetric model including spiral arms \citep{Bobylev2023, Guillaume2025}. These two commonly used potentials capture the main symmetric and spiral perturbations relevant for open clusters \citep{Zeng2025}. The spiral-arm model introduces only moderate, systematic changes, slightly higher eccentricities, marginally larger $Z_{\max}$ values, and perigalactic shifts of up to $\sim$0.5 kpc. In both models, the clusters remain on low-eccentricity thin disk orbits, and the differences between the two potentials are comparable to those arising from $Gaia$ DR3 uncertainties. Thus, our main dynamical conclusions are not sensitive to the choice of Galactic potential.

\begin{table}
\centering
\scriptsize
\renewcommand{\arraystretch}{0.75}
\caption{The evolving, kinematical, and dynamical parameters.}
\begin{tabular}{ll|ccccc}
\toprule
&\textbf{Parameter} & \textbf{NGC 663} & \textbf{NGC 2301} & \textbf{NGC 2384a} & \textbf{NGC 2384b} & \textbf{NGC 7510} \\
\hline\hline
\multicolumn{7}{c}{\textbf{Evolving Parameters}} \\
\midrule\
&$R_h$ (pc) & 2.27 $\pm$ 0.66 & 1.03 $\pm$ 0.10 & 4.18 $\pm$ 0.50 & 4.18 $\pm$ 0.50 & 1.79 $\pm$ 0.07 \\
&$T_{\rm relax}$ (Myr) & 28 & 11 & 45 & 69 & 15 \\
&$\tau$ & 0.54 & 10 & 0.22 & 2.03 & 0.67\\
&$\tau_{\rm ev}$ (Myr) &2800 & 1100 & 4500 & 6900 & 1500 \\
\hline\hline
\multicolumn{7}{c}{\textbf{Kinematical Parameters}} \\
\midrule
&$V_{\rm r}$ (km s$^{-1}$) &  -65.00 $\pm$ 0.87 & 25.89 $\pm$ 0.12 & 55.80 $\pm$ 3.76 & 71.58 $\pm$ 0.17 & -43.81 $\pm$ 5.08   \\
&$\overline{V_x}$ (km s$^{-1}$) &  -20.24 $\pm$ 4.50 & 0.75 $\pm$ 0.01 & 4.41 $\pm$ 2.10 & 14.58 $\pm$ 3.82 & -5.15 $\pm$ 0.44   \\
&$\overline{V_y}$ (km s$^{-1}$) &  -20.63 $\pm$ 4.54 & 26.83 $\pm$ 5.18 & 73.67 $\pm$ 8.58 & 98.46 $\pm$ 9.92 & -50.94 $\pm$ 7.13   \\
&$\overline{V_z}$ (km s$^{-1}$) &  -59.16 $\pm$ 7.70 & -10.24 $\pm$ 3.20 & 18.89 $\pm$ 4.35 & 25.62 $\pm$ 5.06 & -53.55 $\pm$ 7.32   \\
&$\overline{U}$ (km s$^{-1}$) &  47.82 $\pm$ 6.92 & -18.46 $\pm$ 4.30 & -73.67 $\pm$ 8.58 & -99.10 $\pm$ 9.96 & 70.74 $\pm$ 8.41   \\
&$\overline{V}$ (km s$^{-1}$) &  -45.03 $\pm$ 6.71 & -19.34 $\pm$ 4.40 & -16.66 $\pm$ 4.08 & -17.78 $\pm$ 4.23 & -19.92 $\pm$ 4.63   \\
&$\overline{W}$ (km s$^{-1}$) &  -4.66 $\pm$ 2.16 & -10.49 $\pm$ 3.24 & -9.92 $\pm$ 3.15 & -20.68 $\pm$ 4.55 & -9.37 $\pm$ 3.06   \\
&$S_{\rm \odot}$ (km s$^{-1}$) &  65.85 $\pm$ 8.11 & 28.72 $\pm$ 5.36 & 76.18 $\pm$ 8.73 & 102.78 $\pm$ 10.14 & 74.09 $\pm$ 8.61   \\
&$U_{\rm LSR}$ (km s$^{-1}$) &  60.22$\pm$0.84 &-9.94$\pm$0.33   &-67.34$\pm$8.95  &-65.36$\pm$1.37  &91.97$\pm$10.24   \\
&$V_{\rm LSR}$ (km s$^{-1}$) &  -28.45$\pm$1.79 &-4.59$\pm$0.64  &-0.58$\pm$1.72   &-21.08$\pm$3.40  &-0.92$\pm$9.01   \\
&$W_{\rm LSR}$ (km s$^{-1}$) &  -0.69$\pm$0.98 &-2.83$\pm$0.72   &-3.53$\pm$0.35  &-4.66$\pm$0.44    &-5.41$\pm$1.80   \\
&$S_{\rm LSR}$ (km s$^{-1}$) &66.60$\pm$2.21   &11.32$\pm$1.02   &67.43$\pm$9.13  &68.84$\pm$6.17    &92.13$\pm$13.77\\

\hline\hline
\multicolumn{7}{c}{\textbf{Dynamical Parameters}} \\
\midrule
&$A_o~(^\circ)$ &  -134.45 $\pm$ 0.10 & 88.41 $\pm$ 0.11 & 86.57 $\pm$ 0.12 & 81.57 $\pm$ 0.13 & -95.77 $\pm$ 0.13   \\
&$D_o~(^\circ)$ &  -63.81 $\pm$ 0.13 & -20.88 $\pm$ 0.22 & 14.36 $\pm$ 0.18 & 14.43 $\pm$ 0.18 & -46.29 $\pm$ 0.26   \\
&$l_A~(^\circ)$ &  43.28 & -46.35 & -12.75 & -10.17 & 15.73   \\
&$b_A~(^\circ)$ &  4.06 & 21.43 & 7.49 & 11.61 & 7.26   \\
\hline
&$T_{\rm p}$ (Myr) & 258 $\pm$ 3.00 & 242 $\pm$ 1.00 & 309 $\pm$ 3.00 & 288 $\pm$ 5.00 & 325 $\pm$ 4.00 \\
\multirow{5}{*}{\rotatebox{90}{\underline{axisymmetric}}}
&$Z_{\rm max}$ (kpc) & 0.029 $\pm$ 0.002  & 0.042 $\pm$ 0.002   & 0.122 $\pm$ 0.009   & 0.142 $\pm$ 0.007   & 0.108 $\pm$ 0.008 \\
&$R_{\rm a}$ (kpc) & 10.159 $\pm$ 0.04    & 8.81 $\pm$ 0.02    & 11.48 $\pm$ 0.12    & 10.47 $\pm$ 0.09    & 12.54 $\pm$ 0.10 \\
&$R_{\rm p}$ (kpc) & 8.00 $\pm$ 0.15      & 8.36 $\pm$ 0.01   & 9.92 $\pm$ 0.06     & 9.63 $\pm$ 0.23     & 9.79 $\pm$ 0.18 \\
&$R_{\rm m}$ (kpc) & 9.08 $\pm$ 0.09     & 8.56 $\pm$ 0.01   & 10.70 $\pm$ 0.09    & 10.05 $\pm$ 0.16    & 11.16 $\pm$ 0.24 \\
&$e$               & 0.119 $\pm$ 0.001   & 0.026 $\pm$ 0.001   & 0.073 $\pm$ 0.001   & 0.042 $\pm$ 0.001   & 0.123 $\pm$ 0.001 \\
\hline
\multirow{5}{*}{\rotatebox{90}{\underline{non-axisymmetric}}} 
& $Z_{\rm max}$ (kpc) & 0.031 $\pm$ 0.003  & 0.043 $\pm$ 0.001  & 0.124 $\pm$ 0.008   & 0.143 $\pm$ 0.006   & 0.110 $\pm$ 0.006 \\
& $R_{\rm a}$ (kpc) & 10.261 $\pm$ 0.05   & 9.16 $\pm$ 0.03    & 11.82 $\pm$ 0.14   & 10.53 $\pm$ 0.08     & 12.66 $\pm$ 0.12 \\
& $R_{\rm p}$ (kpc) & 7.93 $\pm$ 0.12     & 7.91 $\pm$ 0.02   & 9.54 $\pm$ 0.08    & 9.59 $\pm$ 0.20      & 9.74 $\pm$ 0.18 \\
& $R_{\rm m}$ (kpc) & 9.09 $\pm$ 0.09    & 8.53 $\pm$ 0.02   & 10.68 $\pm$ 0.11   & 10.06 $\pm$ 0.14    & 11.20 $\pm$ 0.15 \\
& $e$               & 0.129 $\pm$ 0.001   & 0.073 $\pm$ 0.003 & 0.107 $\pm$ 0.004   & 0.046 $\pm$ 0.001   & 0.130 $\pm$ 0.001 \\
\bottomrule
\end{tabular}
\label{tab:full_parameters}
\end{table}

\section{Substructures: NGC 2384}
\label{sec:substructure}
One of the key findings of this study is the identification of significant substructures within the region traditionally classified as the open cluster NGC 2384. While previous studies often treated this system as a single stellar group, our analysis of $Gaia$ DR3 astrometry and photometry clearly resolves it into two distinct components, which we designate as NGC 2384a and NGC 2384b. These two groups are not only separated in projection on the sky but also display pronounced differences in their astrometric, photometric, and dynamical properties.

The fundamental parameters summarized in Table~\ref{all_results} indicate that these two subsystems are located at different distances. $Gaia$ DR3 trigonometric parallaxes place NGC 2384a at $0.343 \pm 0.070$ mas ($2.92 \pm 0.60$ kpc) and NGC 2384b at $0.296 \pm 0.068$ mas ($3.38 \pm 0.54$ kpc). Perhaps even more striking is the difference in age. Isochrone fitting reveals that NGC 2384a is a very young system, with $\log t = 7.00^{+0.12}_{-0.11}$ (approximately 10 Myr), similar to other young star-forming regions in our sample, such as NGC 663 and NGC 7510. In contrast, NGC 2384b is a considerably more evolved cluster, with $\log t = 8.15^{+0.17}_{-0.18}$ (around 140 Myr), placing it at a similar evolutionary stage to NGC 2301.

This age difference is directly reflected in their dynamical states. The relaxation time of NGC 2384a ($T_{\rm relax} = 45$ Myr) implies an evolutionary parameter of $\tau = \mathrm{age}/T_{\rm relax} \approx 0.22$, indicating that the cluster is dynamically unevolved. In contrast, NGC 2384b ($T_{\rm relax} = 69$ Myr) is dynamically relaxed with $\tau \approx 2.1$, consistent with its older age. These findings strongly suggest that the two groups, differing both in distance and age, are not a single, coeval cluster.

Their apparent proximity on the sky raises the question of whether NGC 2384a and NGC 2384b could form a physically bound binary OCs. The formation and evolution of binary or multiple clusters is important into the initial conditions of star formation and tidal interactions \citep[e.g.,][]{Delafauntemarcos2009, Palma2024}. To explore this possibility and independently verify the distances and stellar parameters of the two components, we carried out a detailed Spectral Energy Distribution (SED) analysis.

\subsection{Spectral Energy Distribution Analysis}
\label{sec:SED}

We performed a detailed SED analysis for selected members of NGC 2384a and NGC 2384b using the {\sc ARIADNE} Python package, a Bayesian framework for modeling stellar photometry \citep{Vines2022}. We constructed SEDs by compiling photometry from optical to infrared wavelengths. This includes $Gaia$~DR3 $G$, $G_{\rm BP}$, and $G_{\rm RP}$ bands \citep{GaiaDR3}, 2MASS near-infrared data ($J$, $H$, $K_s$; \citealt{Skrutskie2006}), and mid-infrared fluxes from {\it WISE} ($W1$–$W4$; \citealt{Wright2010}). When available, Pan-STARRS and APASS measurements were added to extend the wavelength coverage, enhancing the precision of derived stellar parameters.

\begin{figure}[h]
\centering
\includegraphics[width=0.85\linewidth]{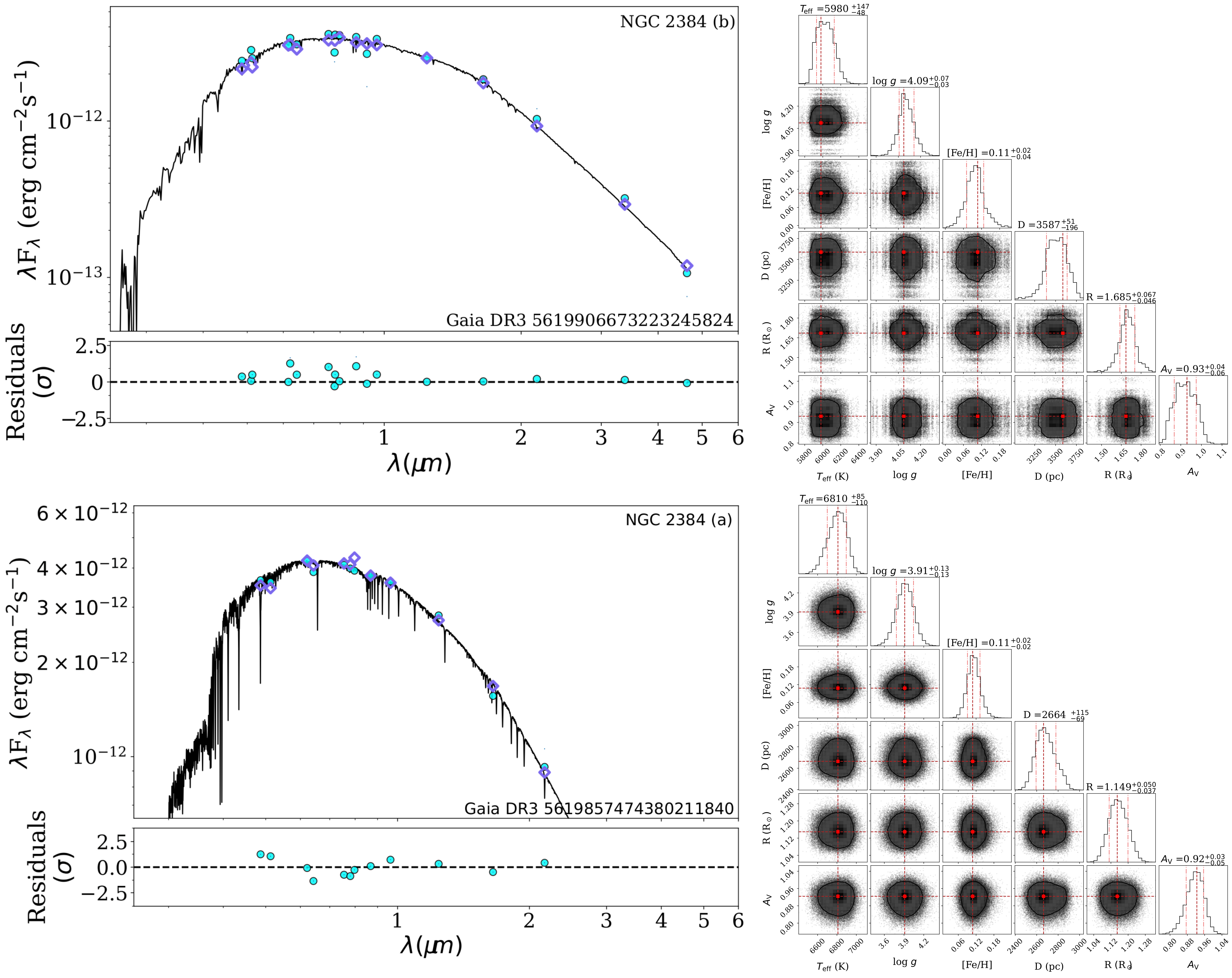}
\caption{Examples of SED fits for two member stars of the open clusters NGC 2384a and NGC 2384b are shown in the left panels, while the right panels illustrate the histograms and distributions of the corresponding astrophysical parameters obtained from the fitting.
\label{Fig: sed_corner}}
\end{figure}

Parameter estimation was conducted with a nested sampling algorithm implemented in {\it dynesty}, yielding effective temperatures ($T_{\rm eff}$), surface gravities ($\log g$), metallicities ([Fe/H]), $V$-band extinctions ($A_{\rm V}$), and stellar radii. Distances were constrained using $Gaia$ EDR3 values from \citet{BailerJones2021}, and line-of-sight extinction was estimated from the dust maps of \citet{Schlafly2011} to ensure consistent color excess corrections. Final stellar parameters were determined via Bayesian model averaging to minimize biases from any single atmospheric model \citep{Alzahrani2025b, Cinar_2024}. SED fitting was carried out only for NGC 2384a and NGC 2384b due to the possibility that these two systems could form a binary cluster. Quality cuts were applied (RUWE $>$ 1.4 and $G > 17$ mag), resulting in 105 NGC 2384a stars and 175 NGC 2384b stars included in the analysis. Figure~\ref{Fig: sed_corner} illustrates representative examples of the SED fits and the corresponding corner plots for one star in each subgroup. The left panels show the observed photometric fluxes alongside the best-fitting synthetic spectra, while the lower panels present the residuals, demonstrating the overall quality of the fits. The right panels display the joint posterior distributions of the inferred stellar parameters, confirming well-constrained solutions for $T_{\rm eff}$, $\log g$, [Fe/H], distance, radius, and $A_{\rm V}$ for both NGC~2384a and NGC~2384b.

\begin{figure}[h]
\centering
\includegraphics[width=0.75\linewidth]{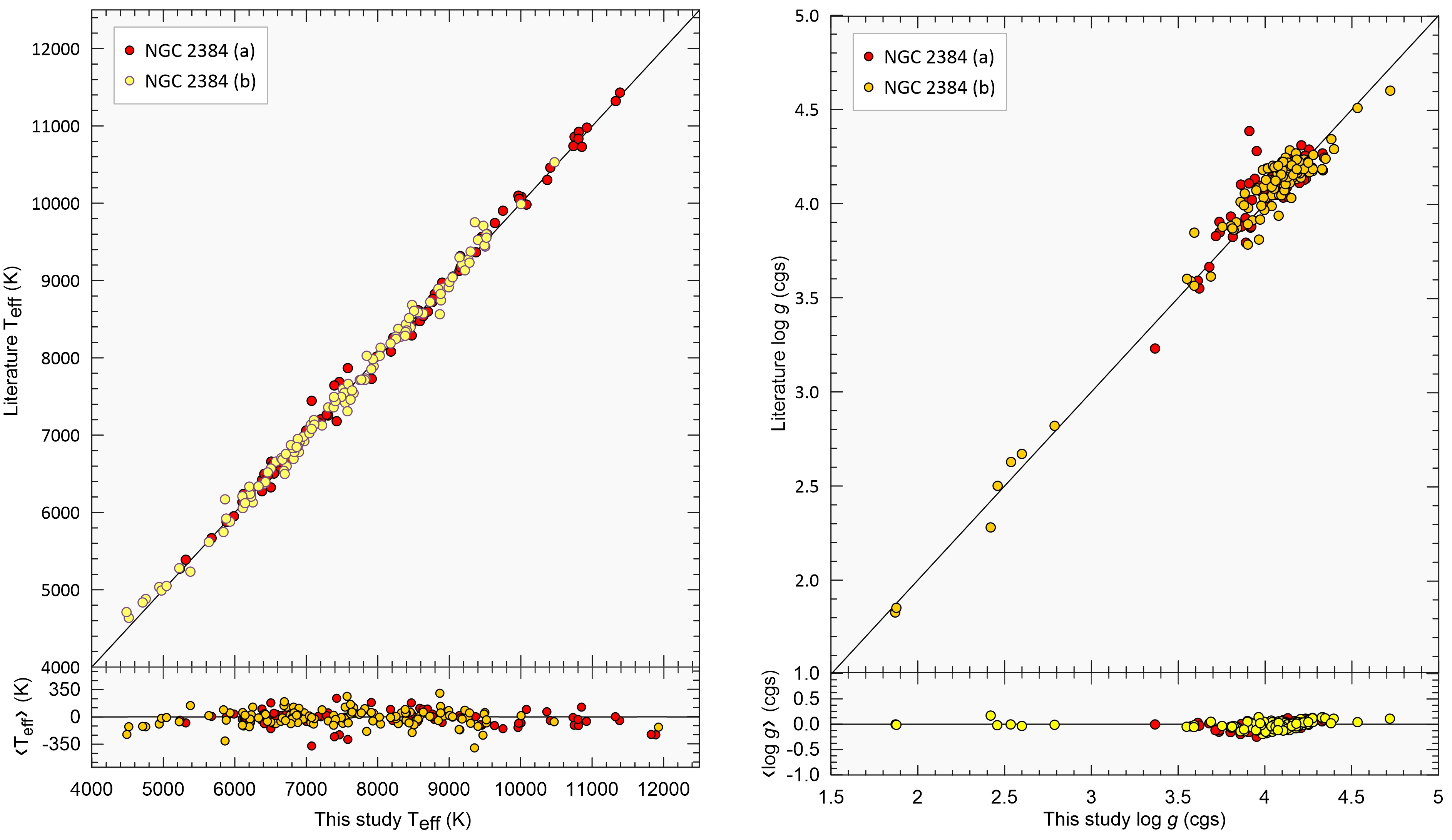}
\caption{Comparison of this study’s atmospheric parameters with literature values for stars in NGC 2384a (red) and NGC 2384b (yellow). The left panel shows $T_{\rm eff}$, the right panel shows $\log g$. The black diagonal marks equal values, and the lower panels display the corresponding residuals.
\label{Fig: lit_sed}}
\end{figure}

We cross-matched stars in NGC 2384a and NGC 2384b with the Gaia DR3 XP stellar-parameter catalogue of \citet{Kordopatis2024}, which provides SHBoost-based estimates for millions of sources. Only entries with reliable XP solutions (\texttt{flag = 0} for $T_{\rm eff}$ and $\log g$) were retained. After applying these criteria, 101 stars in NGC 2384a and 127 stars in NGC 2384b had high-quality XP counterparts, enabling a direct comparison with our SED-derived parameters. As shown in Figure~\ref{Fig: lit_sed}, the agreement is generally good: the median difference in $T_{\rm eff}$ is 7 K with a standard deviation of 97 K, while for $\log g$ the median is 0.02 dex with a standard deviation of 0.07 dex.

\begin{figure}[h]
\centering
\includegraphics[width=1\linewidth]{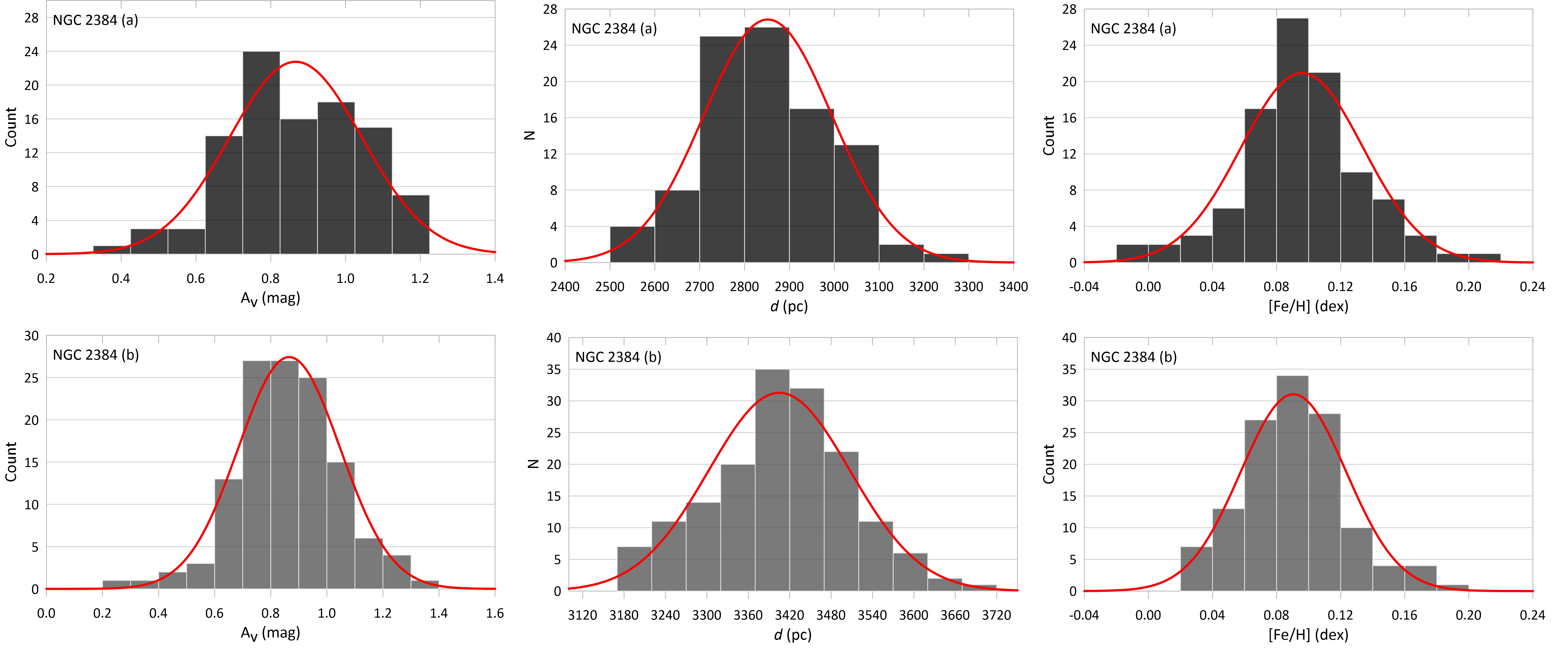}
\caption{Histograms of the member star $V$-band extinctions ($A_{\rm V}$), distances ($d$), and metallicities ([Fe/H]) obtained from the SED analysis, with Gaussian fits overlaid in red in each panel.
\label{Fig: sed_histograms}}
\end{figure}

To characterise the physical distinction between the two stellar groups associated with NGC~2384, we examined the SED-derived parameter distributions for each component. Figure~\ref{Fig: sed_histograms} presents the median values of the fitted stellar parameters, from which the analysis yields distances of $d = 2851 \pm 143$\,pc for NGC~2384a and $d = 3405 \pm 103$\,pc for NGC~2384b, indicating a pronounced distance offset corresponding to a line-of-sight separation of 554\,pc. Despite this clear difference in heliocentric distance, both components display nearly identical extinction values, with $A_V = 0.86 \pm 0.18$ mag for NGC~2384a and $A_V = 0.87 \pm 0.17$ mag for group NGC~2384b, suggesting similar dust conditions along the sightline. Their metallicities are likewise consistent, measured as $\mathrm{[Fe/H]} = 0.10 \pm 0.04$ dex for NGC~2384a and $\mathrm{[Fe/H]} = 0.09 \pm 0.03$ dex for NGC~2384b. These results show that while the two populations share comparable chemical and extinction properties, they are located at significantly different distances along the same line of sight.

\begin{figure}[ht]
    \centering
    \includegraphics[width=0.55\linewidth]{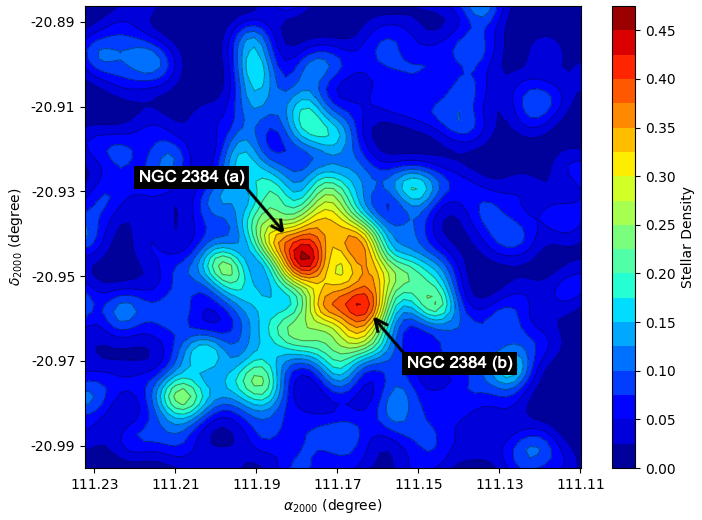}
    \caption{Smoothed stellar density map of NGC 2384 based on the Gaia-selected members. The contour levels represent the stellar surface density obtained from a Gaussian–smoothed 2D histogram. Two distinct overdensities are evident: the northern concentration, labeled NGC 2384 (a), and the southeastern concentration, labeled NGC 2384 (b). The positions of both substructures are indicated by arrows. }
    \label{fig:2384_density}
\end{figure}

The stellar density map (Figure \ref{fig:2384_density}) shows that NGC 2384 consists of two distinct overdensities separated by several arcminutes. Although such a double-peaked morphology may resemble a binary or subclustered configuration, our analysis demonstrates that this structure is purely a projection effect. The two components exhibit clearly different distances, proper-motion centers, and ages, indicating that they do not share a common spatial, kinematic, or evolutionary origin. These diagnostics collectively confirm that NGC 2384(a) and NGC 2384(b) do not form a physical binary cluster but instead constitute an optical pair. In particular, the physical separation between the two components strongly rules out any gravitationally bound configuration. Previous studies show that primordial or tidally captured binary clusters rarely exceed separations of 20–30 pc \citep{Delafauntemarcos2009}. Moreover, the large age difference (10 Myr vs. 140 Myr) is incompatible with the coeval nature required for a primordial binary cluster. We therefore conclude that NGC 2384(a) and NGC 2384(b) are merely an optical pair: two unrelated star clusters that appear close together in projection but are distinct in both distance and evolutionary stage.

\section{Summary and Conclusions}
\label{Summary and conclusion}

In this study, we presented a comprehensive photometric and astrometric investigation of four young open clusters (NGC 663, NGC 2301, NGC 2384, and NGC 7510) located near the Galactic mid-plane, based on $Gaia$ DR3 data. Our analysis demonstrates that the derived fundamental parameters distances, ages, reddening values, mass functions, and dynamical properties are in good agreement with, yet more refined than, earlier literature estimates. Cluster membership was reassessed using the {\sc UPMASK} method, resulting in robust samples of 1498, 575, 337, 622, and 723 stars for the respective systems, each with a membership probability $\geq 50\%$. The mean trigonometric parallaxes derived from \textit{Gaia} DR3 were 0.346 $\pm$ 0.054 mas (2.89 $\pm$ 0.06 kpc) for NGC 663, 1.133 $\pm$ 0.113 mas (0.88 $\pm$ 0.02 kpc) for NGC 2301, 0.343 $\pm$ 0.070 mas (2.92 $\pm$ 0.60 kpc) for NGC 2384a, 0.296 $\pm$ 0.068 mas (3.38 $\pm$ 0.54 kpc) for NGC 2384b, and 0.317 $\pm$ 0.056 mas (3.16 $\pm$ 0.56 kpc) for NGC 7510, yielding distance estimates that are fully consistent with the photometric determinations.  

The ages of the OCs were constrained by isochrone fitting with solar-metallicity models, providing 15 $\pm$ 5 Myr for NGC 663, 110 $\pm$ 10 Myr for NGC 2301, 10 $\pm$ 5 Myr for NGC 2384a, 140 $\pm$ 20 Myr for NGC 2384b and 10 $\pm$ 5 Myr for NGC 7510. These results clearly separate the clusters into two distinct evolutionary groups: extremely young systems still in their early dynamical phases (NGC 663, NGC 2384a, and NGC 7510) and more evolved clusters approaching or surpassing 100 Myr (NGC 2301 and NGC 2384b). The reddening values, $E(B-V) = 0.891$ mag for NGC 663, 0.093 mag for NGC 2301, 0.411 mag for NGC 2384a, 0.341 mag for NGC 2384b, and 1.085 mag for NGC 7510, reflect the significant extinction associated with clusters embedded in or near the Galactic plane, particularly for NGC 663 and NGC 7510.  

The stellar population analysis further revealed slopes of the mass function of $\alpha = 2.23 \pm 0.07$ (NGC 663), $2.00 \pm 0.03$ (NGC 2301), $2.12 \pm 0.07$ (NGC 2384a), $2.26 \pm 0.07$ (NGC 2384b) and $2.08 \pm 0.07$ (NGC 7510), all in close agreement with the canonical Salpeter value ($\alpha = 2.35$). The estimated total stellar content of the clusters spans nearly an order of magnitude, from $\sim 486\,M_\odot$ for NGC 2301 to $\sim 3584\,M_\odot$ for NGC 663, with NGC 7510 ($1863\,M_\odot$) and the two NGC 2384 substructures ($725\,M_\odot$ and $832\,M_\odot$) occupying intermediate positions. Such diversity in total mass emphasizes the heterogeneous nature of young stellar aggregates in the Galactic disk.  

The relaxation times, mean space velocity components, convergent point (CP) solutions, and solar motion corrected velocities derived in this study allow a more detailed interpretation of the present dynamical states of the clusters. The values listed in Table~\ref{tab:full_parameters} indicate that the internal velocity dispersions remain small and that the $(U, V, W)$ components fall within the ranges typically reported for young OCs. This behaviour is consistent with recent \textit{Gaia}-based studies, which show that the dominant kinematic signatures of young clusters largely reflect the motions inherited from their parent molecular clouds \citep{Soubiran_2018, Cantat_2020}.

The relaxation times further reveal that most clusters are still in a dynamically immature phase. Except for NGC~2301 and NGC~2384b, the ratio $\omega = \mathrm{age}/T_{\mathrm{relax}}$ remains below unity, indicating that full energy equipartition through two–body interactions has not yet been achieved. This result is in agreement with recent findings suggesting that young OCs often retain primordial spatial and kinematic substructures for tens to hundreds of Myr \citep{Kuhn2019, Pang2021}. The presence of such substructures in NGC~2384, as identified in this work, provides additional evidence that this cluster has not undergone complete dynamical mixing. Consequently, the internal kinematics of these systems still bear the imprint of their natal conditions rather than long–term dynamical evolution.

The mean Galactic space velocities reinforce this interpretation. The small vertical velocity components ($W \lesssim 10~\mathrm{km\,s^{-1}}$) demonstrate that the clusters remain tightly confined to the Galactic mid–plane and have not experienced significant vertical heating. Such behaviour is typical of young systems located near star–forming regions in the Local and Perseus Arms, and is consistent with the trends reported in the literature \citep{KounkelCovey2019, Zari2019}. In several clusters, most notably NGC~663 and NGC~7510, the negative $V$ values suggest a mild lag with respect to the Local Standard of Rest, possibly produced by spiral–arm streaming motions, local turbulence, or residual gas expulsion during the early stages of cluster evolution.

The convergent point (CP) analysis provides further evidence for coherent bulk motion within the clusters. The CP coordinates exhibit compact groupings in the AD diagrams, indicating that member stars share a well–defined direction of motion. This behaviour closely resembles the CP patterns observed in well–studied young clusters such as the Pleiades and $\alpha$~Persei \citep{2012A&A...538A..23G, Elsanhoury2018}. Moreover, clusters with more extensive radial velocity sampling, particularly NGC~2301 and NGC~2384b, display notably lower scatter in their CP solutions, suggesting that their spatial motions are more tightly constrained and dynamically stable.

The comparison of orbital parameters obtained using the axisymmetric {\sc MWPotential2014} and the non-axisymmetric MW + Spiral potentials reveals consistent and quantifiable systematic changes across the four clusters. Incorporation of spiral arm perturbations produces a modest but systematic increase in orbital eccentricity ($e$), rising from $0.026$--$0.123$ in {\sc MWPotential2014} to $0.046$--$0.130$ in the MW + Spiral potentials. In relative terms, the increase in $e$ ranges from $\sim5$--$10\%$ for NGC~663 and NGC~2384b to $\sim40$--$50\%$ for NGC~2384a, and reaches $\sim180\%$ for NGC~2301, indicating that spiral perturbations have their strongest impact on dynamically cold, nearly circular orbits.

Vertical excursions are only slightly enhanced under the MW + Spiral model: $Z_{\max}$ shifts from $0.029$--$0.142$~kpc to $0.031$--$0.143$~kpc, corresponding to relative increases of only $\sim2$--$7\%$, confirming that spiral structure affects radial heating more strongly than vertical motions on Myr timescales. At the same time, the perigalactic distances ($R_{\rm p}$) exhibit systematic inward shifts of $\sim1$--$6\%$ (up to $\simeq 0.45$~kpc), with the largest changes seen in NGC~2301 and NGC~2384a. Mean Galactocentric radii ($R_{\rm m}$) similarly decrease by $\sim0.5$--$3\%$ under the MW + Spiral potential.

These coherent changes across all clusters demonstrate that spiral arm perturbations can slightly modify the clusters' inferred birth radii and radial oscillation amplitudes, consistent with theoretical expectations that spiral structure contributes to radial migration in the Galactic disk \citep{Sellwood2002, MartinezMedina2015}.

The results confirm that NGC 663, NGC 2384a, and NGC 7510 are very young clusters with ages $\le$ 20 Myr, while NGC 2301 and NGC 2384b represent somewhat older populations at ~100–150 Myr. The wide range of ages among the sample provides a valuable opportunity to probe early cluster evolution in the Galactic disk. The membership determination with \textit{Gaia} astrometry allowed us to uncover a significant substructure in NGC 2384, revealing two kinematically distinct stellar groups, one of which (NGC 2384b) is found to be dynamically relaxed, a result not previously reported in the literature.

The mass function slopes of all clusters are consistent with the Salpeter IMF, indicating that there is no strong deviation from the canonical stellar formation processes. However, the large spread in total cluster mass, from ~500 $M_{\odot}$ in NGC 2301 to ~3600 $M_{\odot}$ in NGC 663, highlights the diversity of stellar populations in clusters of similar evolutionary stages. Dynamical relaxation times reveal that only the older clusters (NGC 2301 and NGC 2384b) have reached a relaxed state, while the younger clusters remain dynamically evolving and are still subject to significant environmental influences.

Based on the results of the SED analysis, and following the classification scheme of \citet{Palma2024}, binary and grouped open clusters are categorized as binaries, capture pairs, or optical pairs, with the tidal factor (TF) providing an additional measure of mutual interaction. Our SED-derived distances indicate that NGC 2384a and NGC 2384b are separated by nearly 0.55 kpc along the line of sight. This separation far exceeds the typical range for physical binary clusters, which is generally limited to a few tens of parsecs. \citet{Delafauntemarcos2009} emphasize that primordial or tidally captured binary clusters rarely exhibit separations larger than ~20–30 pc, as larger distances preclude the possibility of gravitational binding. Furthermore, the significant age difference between the two subsystems is inconsistent with the nearly coeval nature expected for primordial binaries. Consequently, despite their apparent proximity in projection, the SED analysis strongly supports the conclusion that NGC 2384a and NGC 2384b do not constitute a true binary cluster, but are instead an optical pair, in agreement with the classification framework of both \citet{Palma2024} and \citet{Delafauntemarcos2009}.

\begin{acknowledgements}
We would like to thank the anonymous referee for their helpful feedback and recommendations, which have contributed meaningfully to the improvement of our manuscript. We also gratefully acknowledge Dr. Olcay Plevne for valuable technical assistance. This study presents results derived from the European Space Agency (ESA) space mission Gaia. The data from Gaia are processed by the Gaia Data Processing and Analysis Consortium (DPAC). Financial support for DPAC is provided by national institutions, primarily those participating in the Gaia Multi-Lateral Agreement (MLA). For additional information, the official Gaia mission website can be accessed at \url{https://www.cosmos.esa.int/gaia}, and the Gaia archive is available at \url{https://archives.esac.esa.int/gaia}. The authors would like to express their gratitude to the Deanship of Scientific Research at Northern Border University, Arar, KSA, for funding this research under project number "NBU-FFR-2025-237-07". 
\end{acknowledgements}

\bibliographystyle{raa}
\bibliography{ms2025-0428}

\end{document}